\documentstyle[aaspp4,12pt]{article}

\def\etal{{\it et al.$\,\,$}}
 
\def\vperp{{V_{\perp}}}

\def\reobs{{\rm ro}}
\def\creobs{{\rm cro}}
\def\ureobs{{\rm uro}}
\def\survey{{\rm su}}
\def\Nsurvey{{N_{\survey}}}
\def\threshsu{{I_{\rm T}^{(\survey)} }}
\def\threshro{{I_{\rm T}^{(\reobs)} }}
\def\Nave{{\langle N \rangle}}
\def\cl{{ {\cal L} }}
\def\kru{ {K_{\rm r,u}} }
\def\krc{ {K_{\rm r,c}} }
\def\krucrit{{ [\kru]_{\rm crit}}}
\def\krccrit{{ [\krc]_{\rm crit}}}
\def\Ntwid{{\tilde{N}_{\rm D}}}
\tighten
\singlespace
\lefthead{Cordes, Lazio, \& Sagan}
\righthead{Intermittency in SETI}
\begin{document}
\title{SCINTILLATION-INDUCED INTERMITTENCY IN SETI}
\author{James M. Cordes\altaffilmark{1,2,3}, 
	T. Joseph W. Lazio\altaffilmark{1,2}, \& 
	Carl Sagan\altaffilmark{1,3,4,5}}
\altaffiltext{1}{Department of Astronomy, Cornell University}
\altaffiltext{2}{National Astronomy \& Ionosphere Center, Cornell 
	University}
\altaffiltext{3}{Center for Radiophysics and Space Research, Cornell 
	University}
\altaffiltext{4}{The Planetary Society, Pasadena, CA 91106}
\altaffiltext{5}{Deceased}

\setcounter{footnote}{0}

\begin{abstract}

We use scattering theory, simulations, and empirical constraints on
interstellar scintillations
to discuss the intermittency
of radio signals from extraterrestrial intelligence (ETI).  
The number of ETI sources in the Galaxy has a direct influence
on the expected  dynamic range of fluxes in a survey, through inverse
square-law effects and, equally importantly,
 by the number of independent statistical trials
made on modulations caused by interstellar scintillations.  
We demonstrate that scintillations are very likely to allow
initial detections of narrowband signals, while making 
redetections extremely improbable, a result that follows
from the skewed, exponential distribution of the modulation.
This conclusion holds for 
relatively distant sources but does not apply to radio SETI toward
nearby stars ($\lesssim 100$ pc).

Recent SETI has found non-repeating, narrowband events
that are largely unexplained.  We consider three models
in order to assess these events and to analyze
large surveys in general:
(I) Radiometer noise fluctuations;
(II) A population of constant Galactic sources 
which undergo deep fading and amplification due 
to interstellar scintillation, consistent with ETI transmissions; and
(III) Real, transient signals (or hardware errors) 
of either terrestrial or
extraterrestrial origin.

We derive likelihood and Bayesian tests of the models for individual
events and globally on entire surveys.  Applying them to The Planetary
Society/Harvard META data, we find that Models~II and III are both
highly preferred to Model~I, but that Models~II and III are about
equally likely. In the context of Model~II, the likelihood analysis
indicates that candidate events above threshold ($\sim 32\sigma$)
are {\em combinations\/}  of large amplitude noise fluctuations
and scintillation gains,  making it highly probable that
events seen once will only very rarely be seen again.
  Ruling out Model~II in favor of Model~III is
difficult --- to do so, many more reobservations (e.g., thousands) are
needed than were conducted in META (hundreds) {\em or\/}  the
reobservation threshold must be much lower than was used in META.
{\it We  cannot, therefore,  rule out the possibility that META 
events are real, intrinsically steady ETI signals.}

Our formalism can be used to analyze any SETI program.
We estimate the number of reobservations required
to rule out Model~II in favor of Model~III,  taking into
account that reobservations made promptly 
sample the same scintillation gain as in the original
detection,  while delayed reobservations  sample 
a decorrelated scintillation modulation.  
The required number is a strong function
of the thresholds used in the original survey and in
reobservations.

We assess optimal methods for applying statistical tests in future
SETI programs that use multiple site and multiple beam observations as
well as single site observations.  We recommend that results be
recorded on many more events than have been made to date.  
In particular, we suggest that surveys use thresholds that are
far below the false-alarm threshold that is usually set to
yield a small number of noise-induced ``detections'' in a
massive survey.  Instead, large numbers of events should be recorded
in order to 
(1) demonstrate that background noise conforms to the 
    distribution expected for it; and 
(2) investigate departures from the expected noise distribution 
    as due to interference or to celestial signals.
In this way, celestial signals can be investigated at levels 
much smaller than the false-alarm threshold.  The threshold
level for archiving candidate intensities and their corresponding
sky positions is best defined in terms of  the recording and 
computational technology  that is available at a cost commensurate
with other survey costs. 

\end{abstract}


\section{INTRODUCTION}\label{sec:intro} 

The commonly anticipated signal in radio searches for
extraterrestrial intelligence (SETI) is a narrowband,
slowly modulated signal of sustained duration.  For
example, bandwidths less than 1 Hz and signal durations
of days and longer are often hypothesized.   
These signal characteristics are closely related to the 
traditional requirement in science for {\em repeatability\/}  of
an experimental or observational result as well as to the
hope that members of one class of ETI 
signal---those that are targeted
explicitly to us---will sustain themselves
long enough to establish repeatability.

By contrast with such an idealized type of signal, it is easy to
imagine other kinds of received signals that depart significantly from
the idealization, particularly with regard to duration and amplitude
stability.  Moreover, several recent searches have found signals that
are narrowband, do not match the characteristics of known, interfering
signals of terrestrial or solar system origin, and yet, do not repeat
when follow-up observations of the relevant sky positions are
performed.  The META project (\cite{hs93}; hereafter HS93) 
found several dozen such
signals at 1420 and 2840~MHz.  Similarly, SERENDIP~III has found
$\gtrsim 100$ transient candidate events at 430~MHz (Bowyer,
Werthimer, \& Donnelly~1994).  An older example of this kind is the
``WOW!'' signal found in the Ohio State SETI program at 1420~MHz
(\cite{dix85}).  For META the characteristic signal duration is no more than
$\sim 1$ minute and all reobservation attempts---ranging from minutes to years
after receipt of the original candidate signal---have failed.
Intense scrutiny of the WOW sky position (e.g., \cite{g94}) has also
failed to redetect a signal.

A feature common to all SETI programs is that, by virtue
of the multidimensional search space 
(direction, frequency, duration, etc.), large numbers of 
independent statistical tests are performed, typically 
$\sim 10^{13}$--$10^{14}$.
The aim is to separate noise, interference, and candidate real signals. 
This winnowing process is based on the rigorously 
known properties of background noise and on an ad hoc expert system
designed to reject terrestrial signals which fail sidereal and
doppler requirements that bona fide celestial signals  must satisfy. 
As a result of the large number of trials, even pure noise can
produce large amplitudes that are statistically highly improbable in a
single trial.
Similarly, unusual interference,  whether of natural or artificial origin,
that fails to be identified by rejection algorithms can be recorded
as candidate signals. 

In this paper, we address signal intermittency in SETI
programs.  Our goals are to 
1) develop the statistical apparatus for
   interpreting transient signals; 
2) assess surveys, particularly META (HS93), 
   that have yielded nonconfirmed events; 
3) prescribe optimal methods for reporting the results of ongoing SETI
   programs; and 
4) improve the search protocols of future SETI programs.

A short list of possible explanations for intermittency includes:
\begin{enumerate}
\item Noise in the receiver electronics, including thermal noise,
cosmic ray induced events, and hardware failures;
\item Radio frequency interference (RFI) whose origin is terrestrial, from Earth
orbit, or from interplanetary spacecraft;
\item Natural, extrinsic  modulation of narrowband 
astrophysical sources such as that caused by
interstellar (radio) scintillation (ISS) or 
gravitational lensing;
\item  Natural, extrinsic  modulation of ETI sources; and
\item Intrinsic  intermittency  at the source of an ETI signal  due to natural
causes, such as planetary rotation or the nature of the transmission 
(e.g., planetary radar) or, for deliberate reasons, to frustrate 
detection and decryption by non-target civilizations.
\end{enumerate}
Causes (2) and (5) are not dissimilar in their transmission properties 
though their sidereal and doppler properties may allow them to  
be distinguished.  In this paper, we use a Bayesian likelihood
analysis as the apparatus for comparing cause (4) with several of the
other, plausible explanations for signal intermittancy.

Our focus on cause (4), with particular emphasis on ISS, is motivated
by two considerations.  First, it is the simplest possible explanation
for the transient behavior of signals that {\em perhaps\/} otherwise
would be steady.  Second, ISS is important at centimeter wavelengths
(\cite{ric90}) that are commonly used in searches for ETI.  Centimetric
wavelengths are often favored because 
background noise levels are minimized  relative to
other wavelength bands and it is argued that 
spectral lines from HI and OH in this ``waterhole'' band
serve as signpost or ``magic'' frequencies at which to transmit
and receive (\cite{oli73}).  Most importantly, however, we find that
ISS {\it by itself} is sufficient to explain the lack of confirmation
of any ETI candidate signals in surveys conducted to date.   


We consider ISS at waterhole frequencies using the framework of Cordes
\& Lazio~(1991) and a recent model for Galactic free electrons by
Taylor \& Cordes~(1993).  We show that the properties of
scintillations expected from a Galactic population of ETI sources are
intimately related to the {\em number\/} of such sources, if
quasi-uniformly distributed, thought to exist in the Galaxy.  This
follows because the time scale and characteristic bandwidth are
strongly dependent on the distance of the source.  In addition, the
maximum increase in amplitude due to scintillation expected for any
source is related to the number of statistical trials made and is
therefore also related to the number of sources in the Galaxy.

In previous papers  
(\cite{cl91}; \cite{cl93}; \cite{lc98},
hereafter CL91, CL93, \& LC98, respectively), 
we have emphasized the need for multiple observations
of the sky in order to combat the effects of interstellar
scintillations.  In particular, we have found that the
optimal number of observations of a specific direction
is a few and that these can restore some of the detection
probability that would be lost otherwise in a single
pass on the sky.   Here, we focus on reobservations
of an initial candidate detection and find a seemingly
contradictory result: namely, that to rule out the 
reality of the candidate signal, it takes hundreds, if
not thousands, of reobservations to do so.

In \S\ref{sec:iss} we summarize the 
effects scintillations impose on celestial signals from compact sources.
In \S\ref{sec:eti_transmit} we
discuss the distribution of a Galactic population of extraterrestrial
civilizations and, for simplicity, we assume that the 
Galactic population is a set of ``standard candle'' transmitters.
In \S\ref{sec:survey_programs} we summarize some recent SETI programs
and the candidates they detected.
%
Basic issues underlying an interpretation of survey results are
outlined qualitatively in \S\ref{sec:basic} while in
\S\ref{sec:obs_events} we introduce the likelihood functions that form
the basis of our quantitative analysis.
In \S\ref{sec:like_events_u} we analyze individual candidate events
assuming that any multiple observations occur at times such that
interstellar scintillations are uncorrelated; correlated
scintillations are treated in \S\ref{sec:like_events_c}.
In \S\ref{sec:surveys} we extend our likelihood analysis to surveys as
a whole.

We extend our analysis to two-station SETI in \S\ref{sec:bistatic}
where we show that many of the challenges in verifying scintillating
signals at a single site persist for simultaneous observations made at
a pair of sites.
We propose an alternative detection scheme in \S\ref{sec:pdfclean} that
uses all of the data in a survey and is much more sensitive than 
a signal-to-noise ratio limited survey.
Our analyses are distilled into recommendations for future SETI 
in \S\ref{sec:future} and in \S\ref{sec:summary} we present our conclusions.

In Appendix~\ref{app:syms} we define symbols used throughout our analysis
and Appendix~\ref{app:stats} discusses scintillating source properties,
including detection probabilities at single and multiple sites.

\section{INTERSTELLAR SCINTILLATIONS (ISS)}\label{sec:iss}

In CL91, CL93, \& LC98,
we have described the 
impact upon source detection procedures resulting from radio waves
propagating through a random, phase-changing medium, such as the
ionized interstellar medium (ISM) or the interplanetary medium (IPM).
The reader is directed to those papers for details on the salient
features of scintillations (also see \cite{ric90}).  Here we summarize
scintillation phenomena that are relevant to our analysis.


\subsection{Brief Summary}\label{sec:scint_summary}

The scattering regime in which we are most interested is that which
produces saturated intensity scintillations.  In this {\em strong
scattering\/} case, intensity variations branch into two 
forms (\cite{ric90}): fast,
diffractive scintillations (DISS) and slow, refractive scintillations
(RISS).  DISS is manifested as 100\% variations of the intensity in
both time and frequency.  The characteristic time and frequency scales
are discussed below.  RISS typically shows 10--20\% variation on times
scales of days to months and is correlated over a wide range of
frequency.  At centimeter wavelengths, most of the Galaxy is seen in
the strong scattering regime.  For sources near the Sun, e.g., $<100$
pc at 1~GHz, weak scintillations occur.  A transition region extending
to about 500~pc at 1~GHz produces scintillations with greater than
100\% intensity variations (c.f. LC98; \cite{ric90}).  Consequently,
in the remainder of this paper, we shall concentrate on the impact of
strong scattering on SETI surveys done to date.  Also, we discuss
only the effects of DISS because modulations from it are orders of
magnitude larger than those from RISS in large-scale SETI.

While our interest here is primarily on {\em intensity variations\/} 
from interstellar scattering, phase fluctuations induced by the medium
may also manifest themselves as {\em Doppler shifts\/}  or {\em
frequency wandering\/} of narrow spectral lines (CL91).  In SETI programs
optimized for finding narrowband signals, follow-up observations on
candidate signals need to account for the possibility that signals
drift or shift in frequency on a variety of time scales.  Drifts may
be caused by interstellar scattering, interplanetary scattering in the
solar wind, and, presumably, in the stellar wind of the transmitting
civilization's host star.  These drifts are much less than the Doppler
shifts caused by the spin or orbital motion of a planet.  Nonetheless,
they must be taken into account when making follow-up observations of
candidate signals because they can still span several to many
frequency bins of a narrow-band spectrometer.



\subsection{Characteristic Time and Frequency Scales}\label{sec:tf}

The intensity of a scintillating source is correlated on
characteristic time and frequency scales: the scintillation time,
$\Delta t_{\rm d}$, and the scintillation bandwidth, $\Delta\nu_{\rm
d}$.
For the vast majority of lines of sight, the scintillation bandwidth exceeds the
anticipated intrinsic widths of deliberately transmitted lines
($\lesssim 1$ Hz).  In addition, broadening of the line by propagation
effects is much less than 1 Hz for most lines of sight.  Therefore,
assuming intrinsic line widths are not substantially less than 1 Hz,
the line shape is essentially unaffected by scattering while the line
amplitude is modulated for most lines of sight (\cite{dh77}; CL91).

When scintillations are saturated and the line shape undergoes
negligible changes, the only astrophysically variable parameter is the
scintillation time scale, $\Delta t_{\rm d} \propto \nu^{1.2} /
\vperp$ where $\vperp$ is a characteristic transverse speed that is a
combination of the velocities of the source, observer, and medium.
The time scale also depends on distance, with more distant
sources having shorter time scale scintillations.  For a medium with
homogeneous statistics, $\Delta t_{\rm d} \propto D^{-0.6}$ 
(Cordes \& Lazio 1991), but the
dependence is more complex for a realistic model of the Galaxy,
as discussed below.
The time scale has been measured to be seconds to hours for pulsars
(\cite{cor86}), whose transverse velocities dominate the relevant
$\vperp$ and are in the range of 30 to 2000 km s$^{-1}$ (\cite{hl93};
\cite{crl93}; \cite{ll94}).  By comparison, the motions of main
sequence stars in the Galaxy and the orbital motions of their
planetary companions are much lower than the average pulsar speed.
Therefore, we expect that an ETI source at the same distance as a
given pulsar will show a characteristic scintillation time scale that
is much longer than that seen from the pulsar, by typically an order
of magnitude.  We point out that differential galactic rotation does
not play a large role because the scattering material also
participates in the rotation (\cite{cor86}).  We also emphasize that,
as the scintillation time is frequency dependent, an assessment of the
role of scintillations must take the observation frequency into
account.


Through modeling of the free electron density in the Galaxy, it is possible
to estimate the scintillation time scale for ETI sources.  The most recent
model for the free electron density and its fluctuations (\cite{tc93})
incorporates all available data on pulsar distances and  dispersion measures,
and radio scattering data on pulsars, Galactic maser sources,
 and extragalactic sources viewed through the ISM.   

Figure~\ref{fig:cls_iss} shows contours of the scintillation time
$\Delta t_{\rm d}$ for directions in the plane of the Galaxy (i.e.,
Galactic latitude $b = 0$).  We have assumed a transverse speed
$\vperp = 10$ km~s$^{-1}$ for all ETI sources and an observation
frequency of 1.42~GHz.  
Given the range of possible orbital motions
and motions in the Galaxy, the effective velocity could vary
by a factor of two or more, either lower or higher.  
It is evident that the scintillation time
scale is of order minutes only for the most distant regions of the
Galaxy.  For most directions, it is typically hours.

The scintillation time scale, therefore, significantly exceeds the
typical time spent on a given sky position in SETI unless all sources
are in the Galactic plane at large distances.  Consequently, one may
consider any putative (intrinsically steady) ETI signal to be constant
over the observation time, though scintillations may have modulated it
significantly upward or downward in apparent strength relative to its
long term mean. Prompt repeat observations of a candidate signal made
sooner than one diffraction time scale after an initial, tentative
detection will ``see'' the same signal strength.  Conversely, repeated
observations made at times when the scintillation modulations are
mutually independent will detect a large range of signal strengths
that follows the distribution of scintillation modulations.  For
saturated scintillations, this distribution is a one-sided exponential
function.

The reader may think that Fig.~\ref{fig:cls_iss} can be used to
dismiss immediately candidate signals in META as being due to RFI,
since they are not redetected immediately after their initial
detections.  This is not the case.  As we show below, candidate
signals can be detected in META-like surveys as a {\em combination\/} 
of noise fluctuations and scintillations of celestial signals.  The
noise decorrelates immediately, so even a persistent source can
disappear into the noise after an initial detection.

Figure~\ref{fig:cls_iss} also shows the scintillation bandwidth,
$\Delta\nu_{\rm d}$, at 1.42~GHz.
The scintillation bandwidth is found to be
much greater than the hypothesized signal bandwidth ($\lesssim 1$ Hz)
of deliberate ETI signals for all lines of sight through the Galaxy.
The diffraction bandwidth scales with frequency as $\Delta\nu_{\rm d}
\propto \nu^{4.4}$.

The reciprocals of the scintillation time scale and bandwidth are also
of potential interest, yielding the bandwidth of {\em spectral broadening\/}
and the {\em pulse-broadening\/} time, respectively, as
$\Delta\nu_{\rm sb} = (2\pi\Delta t_{\rm d})^{-1}$ and
$\tau_{\rm d} = (2\pi \Delta\nu_{\rm d})^{-1}$ (e.g., \cite{l70}; CL91, CL93).
Using values for the scintillation time scale $\Delta t_{\rm d}$ from
Fig.~\ref{fig:cls_iss}, it is clear that spectral
broadening $\ll 1$ Hz at 1.42~GHz.  This confirms our previous
statement that line shapes are essentially unaltered by
scattering if they are intrinsically wider than the
spectral broadening from scattering. 

\subsection{Detection Probabilities}\label{sec:pdfs}

Intensity statistics are discussed in Appendix~\ref{app:stats}.  There
we distinguish variations of the intensity over short time scales,
when the scintillation modulation is constant, from those over long
times when the modulation varies according to its exponential
distribution (in strong scattering).  In the former case the intensity
probability density function (pdf) is that for a phasor of constant
amplitude $gS$ combined with noise, where $g$ is the scintillation
``gain'' and $S$ is the source intensity in the absence of
scintillations.  The pdf $f_I(I; gS)$ is given in Eq.~\ref{eq:pdf},
with $S\to gS$.  In the second case, where $g$ varies over its sample
space, the pdf, $f_{I,\rm scint}(I; S)$, is a simple, one-sided
exponential with mean $\langle I \rangle = S + \langle N \rangle$,
where $S$ is assumed to be constant and $N$ is radiometer noise.
%

Each pdf implies
a {\em detection probability\/} that the intensity
exceeds a specified threshold intensity, $I_{\rm T}$.  We call these
$P_{\rm d}(I_{\rm T}; S)$ and $P_{\rm d,\rm scint}(I_{\rm T}; S)$,
respectively, and define them in
Eq.~(\ref{eq:pd_nonscint})--(\ref{eq:pd_scint}).  When there is no
signal, just noise, $S = 0$, the detection probability is the
false-alarm probability.  In Fig.~\ref{fig:cls_pd} we show detection
probabilities with and without scintillations for the case where the
false-alarm probability is $10^{-12}$, corresponding to
a threshold 
$I_{\rm T}\langle N \rangle = \log 10^{+12} = 27.6$.    
Recent SETI programs do, in fact, achieve 
conformance to exponential statistics up to large thresholds
such as this so long as terrestrial RFI is removed
(cf. Figure 4 of HS93; \cite{lobg95}).

We reiterate a point discussed at length in our earlier papers:
In the absence of scintillations, a source is essentially undetectable
if $S < I_{\rm T}$, i.e., $P(I > I_{\rm T}) \approx 0$.  However, by
the multiplicative nature of the scintillation gain,
otherwise undetectable sources can be modulated above threshold.
Since $g$ is drawn from an exponential distribution, it can reach
large values that result in a detection probability, $P_{\rm
d}(I_{\rm T}, gS)$, significantly in excess of its value in the
absence of scintillations, $P_{\rm d}(I_{\rm T}, S)$.
Modulations which increase the signal strength are not the norm, 
though, and are in fact less likely than the converse.  In contrast to
our earlier papers (CL91, CL93, LC98), where we considered the
probability of detecting a source at least once, in a survey, say, here
we consider the probability of detecting a source at least twice, once
in a survey and then again in a set of reobservations. 

\subsection{Second-order Statistics}\label{sec:order2}

Appendix~\ref{app:stats} also presents results on second-order
statistics that we use in discussing 
correlations in scintillation gain between several observations of the
same source.  Of particular interest is the conditional probability
that a signal is detected in a second observation given that the
intensity had the value $I_1$ in an initial ``detection,'' $P_{\rm
2d}(I_{\rm T} \vert I_1; S, \rho_g[\Delta t])$.  Here $\Delta t$ is
the time interval between observations and $\rho_g$ is the
autocorrelation function of the scintillation gain, normalized to
$\rho_g(0) \equiv 1$.  This function is gaussian-like in form and has
a width equal to the characteristic (diffractive) scintillation time,
$\Delta t_{\rm d}$.


In Fig.~\ref{fig:cls_p2cond} we show the conditional probability
plotted against the correlation coefficient $\rho_g \in [0,1]$ for
several values of the normalized source strength $\zeta \equiv S /
\Nave$, 
threshold $\eta_{\rm T}
\equiv I_{\rm T}/\Nave$, and intensity $\eta_1 \equiv I_1/\Nave$ for an
initial detection. As we discuss in \S\ref{sec:bistatic}, observations
made at two different sites may also be analyzed using
Fig.~\ref{fig:cls_p2cond}.  Like two temporally-spaced observations,
the scintillation gain is partially correlated according to a spatial
correlation function, $\rho_{\rm s}$.


Figure~\ref{fig:cls_p2cond} illustrates the importance of the
combined effects of noise and scintillation.
For values typical of META, 
$\eta_1 = 32$ and $\eta_{\rm T} = 20$, 
the probability $P_{\rm 2d}$ is less than unity
for a nominally detectable signal, $\zeta = 32$,
even for perfectly correlated
scintillation gains. 
If we suppose that
the initial detection, $\eta_1$, occurred because of the
scintillation amplification
of a weaker signal, e.g., $\zeta \le 16$, the chances of {\em not\/} 
detecting the signal during reobservations are even greater.
Combinations of $\zeta$ and $\rho$ can produce 
$P_{\rm 2d} \ll 0.1$
in some cases. 

We will demonstrate that, in large scale surveys involving large
numbers of statistical trials, threshold crossings (initial
``detections'') will occur through the combined effects of noise and
scintillating phasors (in the case where ETI signals exist!).  Because
of this ``scintillation-noise conspiracy,'' 
{\em confirmations\/}  of detected signals
are difficult, owing to the rapid decorrelation of the noise on short
time scales.  Thus, for large surveys with large detection
thresholds, the probability of redetection is small and, therefore, the
absence of confirming redetections is insufficient evidence for
concluding the absence of a possibly steady ETI signal.  Similarly, {\em
simultaneous\/} observations at two sites, where radiometer noise is
uncorrelated, allow detection at one site and nondetection at the
other with large probability.

Qualitatively, noise by itself---without aid from scintillation---
can also assist survey detection while inhibiting redetection.  
But quantitatively,  for thresholds, candidate numbers,
and numbers of reobservation trials like those in META, it is
difficult to explain noise-assisted survey detections
combined with the failure of all reobservations.   Evidently,
scintillations are needed to explain both.   The reason
lies in the difference in shapes of the detection probability
curves in Figure \ref{fig:cls_pd} for intensity ratios
$S/I_{\rm T} \lesssim 1$.   These, in turn, arise
because, even though noise and scintillations both have
one-sided exponential pdf's, they act asymmetrically because
scintillations act multiplicativly on the true signal, while 
noise is additive.

\subsection{Scintillations as a Reality Check on Candidate Sources}
\label{sec:reality}

If a candidate signal is also detected in the
reobservation phase of a SETI program, scintillations can serve
to confirm that the candidate signal is in fact coming from beyond the
solar system (cf. HS93).  The strength of a signal from a
candidate source should vary from observation to observation in a
manner consistent with the exponential distribution, provided the
observations are separated by more than a scintillation time and, of
course, that scintillations are strong.

In CL91, we alluded to another scintillation phenomenon, 
{\em spectral wandering}.  If observed over time periods shorter than the
scintillation time, the centroid of a narrow-band signal will show
random but correlated shifts in frequency in addition to any Doppler shifts.
If the signal is averaged over times much larger than the
scintillation time, the signal will be spectrally broadened, i.e., the
spectral shape of the signal will converge to a Gaussian-like shape with 
width $\Delta\nu_{\rm sb}\sim 1/\Delta t_d$ (CL91).

Expressions are presented in CL91 for $\Delta\nu_{\rm sb}$ as a
function of distance and frequency for both weak and strong
scintillation regimes.  For Galactic sources, we expect frequency
wandering to be roughly the reciprocal of the scintillation time
scale.  Reference to Fig.~\ref{fig:cls_iss} shows that wandering will
be at the level of $10^{-4}$ to 0.1 Hz.  There are some directions for
which the level of scattering is underestimated by the model used to
construct Fig.~\ref{fig:cls_iss}.  For these directions (such as
sources in or beyond the Galactic center), the spectral wandering can
exceed 0.1 Hz.

\section{GALACTIC ETI TRANSMITTERS} \label{sec:eti_transmit}

In this section, we show that the number of ETI transmitters in the
Galaxy influences the expected distribution of source
fluxes and determines the scintillation properties of the sources as
manifested in a particular survey.

\subsection{Demography}\label{sec:demography}

Suppose the number of transmitting civilizations in the Galaxy is
$N_{\rm D}$ as estimated, for instance, from the Drake equation
(\cite{ss66}).  The number density of sources is
\begin{equation}
n_{\rm D} = \frac{N_{\rm D}}{2\pi R_{\rm G}^2 H_{\rm G}},
\label{eq:ndense}
\end{equation}
assuming that all civilizations reside in a disk of
radius $R_{\rm G}$ and thickness $2H_{\rm G}$.

The volume through the Galactic disk sampled by a radio telescope of
beam solid angle $\Omega_{\rm b}$ is $V_{\rm b} = \Omega_{\rm b}
D^3/3$, where $D$ is the characteristic distance through the disk, $D
\approx {\rm min}(H_{\rm G}/\sin\vert b\vert, R_{\rm G})$, where $b$
is the Galactic latitude.  $D$ is direction-dependent, but
observations in the plane will yield $D\sim R_{\rm G}$.  The number of
sources in a typical beam is
\begin{eqnarray}
N_{\rm b}
 = n_{\rm D} V_{\rm b} \sim  10^{-2.7}
N_{\rm D} \left (\frac{D}{R_{\rm G}} \right )^3 \left (\frac{R_{\rm G}
/ H_{\rm G}}{150} \right ) \left (\frac{\theta_{\rm b}}{1\deg}
\right)^2,
\label{eq:n_in_beam}
\end{eqnarray}
where $\theta_{\rm b}$ is the one-dimensional beam width (FWHM) and we
assume $R_{\rm G}/H_{\rm G} = 150$.  If there are more than about
$10^3$ sources in the Galaxy, every telescope beam (of
$1\deg$ size) will contain  at least one source, on average.
It is convenient to rewrite $N_{\rm b}$ as
\begin{equation}
N_{\rm b} = \left ( \frac{D}{D_{\rm mfp}} \right )^3
\label{eq:n_in_beam_mfp}
\end{equation}
where the ``mean free path'' for encountering a source in the beam is
\begin{equation}
D_{\rm mfp} = \left ( \frac{3}{n_{\rm D}\Omega_{\rm b}} \right )^{1/3} 
            \approx 
  R_{\rm G} \left(\frac{501}{ N_{\rm D} } \right )^{1/3}
            \left (\frac{150}{R_{\rm G} / H_{\rm G}} \right )^{1/3} 
            \left (\frac{1\deg}{\theta_{\rm b}} \right )^{2/3}
\label{eq:mfp}
\end{equation}

The typical distance between sources is
%
%
\begin{equation}
\ell_{\rm D} \sim \cases { 
H_{\rm G} 
\left (\frac{2}{3} \right )^{1/2}
\left (\frac{\displaystyle\tilde{N}_{\rm D}}{\displaystyle N_{\rm D}} \right )^{1/2}
\left [ 1 + \frac
       {\displaystyle N_{\rm D}}
       {\displaystyle 2\tilde{N}_{\rm D}} \right ]^{1/2} & $N_{\rm D} \le \Ntwid$; \cr  
\cr
H_{\rm G} \left (\frac{\displaystyle \Ntwid}{\displaystyle N_{\rm D}} \right )^{1/3}
      & $N_{\rm D} \ge \Ntwid$,}
\label{eq:separation}
\end{equation}
where $\Ntwid$ is the number of sources such that $\ell_{\rm D}$ equals one
scale height of the Galactic disk ($H_{\rm G}$):
\begin{equation}
\Ntwid = \frac{3}{2} \left ( \frac{R_{\rm G}}{H_{\rm G}} \right )^2
       = 10^{4.5} \left ( \frac{R_{\rm G}/H}{150} \right )^2.
\label{eq:ntwid}
\end{equation}
Analogous expressions can be derived for a spherical distribution of 
ET sources.  The concentration of META candidates toward the Galactic
plane motivates our focus on a disk population, however. 

Finally, although we assume the strong scattering regime throughout
this paper, there may be sources within the weak or transition
scattering regimes if the number of civilizations in the Galaxy
exceeds $10^5$ and $10^3$, respectively.  
These numbers follow from
our estimates in \S\ref{sec:scint_summary} for the distances
to which weak or transition level ISS occurs and matching these to
Eq.~\ref{eq:separation}  and solving for $N_{\rm D}$. 
Also, the population of Galactic
ET civilizations can be as large as $10^5$ depending on what fraction
of the unaccounted-for events in META are, in fact, due to ETI sources
(HS93).

\subsection{Source Strengths of Standard Candle Transmitters} 
\label{sec:candles}

A Galactic population of transmitters, each radiating
with effective isotropic radiation power (EIRP) ${\cal P}$, 
will show maximum and minimum fluxes,
\begin{eqnarray}
S_1 &=& {\cal P}/4\pi\ell_{\rm min}^2 \\
S_0 &=& {\cal P}/4\pi\ell_{\rm max}^2 
\label{eq:fluxes}
\end{eqnarray}
from the nearest and furthest sources, respectively.
\footnote{ 
For simplicity, we ignore transmitter beaming in
our analysis.   Beaming away from us clearly extends
the minimum flux $S_0$ to zero.  The net effect, of
course, is to decrease the probability of detecting
strong sources when there is a fixed number of
transmitters in the Galaxy.   
}
Within these bounds, the pdf for $S$ is
\begin{equation}
f_S(S) 
 = \left( \frac{\alpha - 1}{S_0} \right ) \left[ 1 -
(S_0/S_1)^{\alpha - 1} \right ]^{-1} \left( \frac{S}{S_0} \right
)^{-\alpha}.
\label{eq:pdf_candles}
\end{equation} 
Disk and spherical populations are described by $\alpha = 2$ and
$\alpha = 5/2$, respectively, as is well known in studies of natural
radio sources (\cite{sch74}; \cite{con74}) and $\gamma$-ray sources
(\cite{was92}).  

Taking $\ell_{\rm min} \sim \ell_{\rm D}$, 
and $\ell_{\rm max} \sim
R_{\rm G}$, we find that $S_1 / S_0 \sim N_{\rm D}$  
for $N_{\rm D} \le \Ntwid \sim 10^{4.5}$.   
In the large civilization limit ($N_{\rm D} \gtrsim \Ntwid$), the scaling is 
$S_1/S_0 \sim (R_{\rm G}/H_{\rm G})^2\, (N_{\rm D}/\Ntwid)^{2/3}\sim 10^{4.4}\,(N_{\rm D} / \Ntwid)^{2/3}$ 
(c.f. Eq.~\ref{eq:separation}).
Depending on the abundance
of transmitting civilizations in the Galaxy, the dynamic range of
source fluxes may be modest or extremely large.

In reality, radio transmission powers from Galactic civilizations will
be described by a luminosity function for ${\cal P}$ that incorporates
technological and sociological evolution.  Since we do not know the
form of the luminosity function, we will make use of the ``standard candle''
form of Eq.~\ref{eq:pdf_candles} in the remainder of the paper,
keeping in mind that it is essentially a place holder for what may be
a quite different distribution of source strengths.

\section{SURVEY PROGRAMS}\label{sec:survey_programs}

In this section we summarize current and recent survey programs; see
also Table~\ref{tab:surveys}.

\subsection{The META Program}\label{sec:programs}

Horowitz \& Sagan~(1993) summarize
the META Survey that uses a $2^{23}$ point Fourier transform
spectrometer to analyze a total bandwidth of 400 kHz (0.05 Hz
resolution).  Two observing frequencies were analyzed: 1420~MHz in the
``waterhole,'' and its harmonic at 2840~MHz.  Spectra were obtained 
in 20 s data spans at
each sky position after adjusting for the Doppler shift from the
Earth's motion relative to each of three reference frames: the local
standard of rest, the Galactic barycenter, and the cosmic microwave
background.  Over a five-year period, about $6\times10^{13}$
combinations of spectral channels and sky positions were investigated.
Most of these, of course, yielded amplitudes consistent with the
expected exponential noise.  After removing obvious, terrestrial
interference and setting a threshold such that few, noise-only events
would exceed it, 74 events remained.  Of these, half could be rejected
through further investigation as being due to terrestrial
interference, cosmic ray events in the semiconductors of the
spectrometer, and other processor errors.  

The remaining 37 candidates
include 14 at 1420~MHz and 23 at 2840~MHz with approximately equal
numbers of candidates in each of the three reference frames.
Of these 37 events, 26 are formally below the false-alarm threshold
($\sigma \ln 6\times 10^{13} = 31.7\sigma$) defined so that one
event should exceed the threshold in the entire survey.   HS93 comment
that the 26 events below 31.7$\sigma$ are consistent with the
exponential distribution expected from pure noise.  Consequently,
in our analysis, we restrict our attention to the 11 events in
Table 2 of HS93 with amplitudes $> 31.7\sigma$ and especially
to those 9 events larger than $33\sigma$. 

When a candidate detection occurred, the real-time data acquisition
system made reobservations of the sky position in the relevant
reference frame beginning about 40 s after the initial detection and
ending about 3 min afterward, when the source had drifted out of the
primary beam of the telescope (when used as a transit instrument).  It
is notable that {\em none of the candidate positions showed evidence
for a signal in any of the reobservations, either within minutes after
the initial detection, or days, months, or years later}.  The
effective detection threshold in reobservations was significantly
lower than in the main survey observations, being typically 20 times
the mean noise level.


If all 37 META events that survive culling are due to ETI, there
must be $> 10^5$ transmitting sites at 1420 and 2840~MHz in the Galaxy
(HS93), based on a simple scaling of META's ``duty cycle'' in scanning
the sky.

\subsection{The META~II Program}\label{sec:metaii}

The META~II system uses back-end hardware identical to that of
META on a 30~m antenna in Argentina (\cite{col93}).  The observation
frequency is 1.42~GHz.  Like META, META~II finds extrastatistical
events that are not redetected when observations are made at the
appropriate sky positions.  Reobservations are made both promptly and
with substantial delay of a day or more.  For about 25\% of the time
since 1993 November, observations have been made simultaneously with
META and META~II in the overlapping sky region of 
$-30\deg \le \delta \le -10\deg$.

\subsection{The SERENDIP~III Program}\label{sec:serendip}

The SERENDIP~III program (\cite{bow94}) has been operated at the
Arecibo Observatory since 1992 April in a ``piggyback'' mode whereby
pointing of the antenna is under control of other investigators, thus
yielding semi-random but complete coverage of the Arecibo sky.  In
this program, a 12~MHz bandpass centered at 0.43~GHz is analyzed in
2.5~MHz wide subbands.  An FFT spectrometer obtains 4M point spectra
in 1.7 s at each of 5 subbands for a channel width of 0.6 Hz.  The
program has covered 89\% of the observable sky at Arecibo (28\% of the
entire sky) at least once and about 18\% of the observable sky has
been observed at least 5 times.  The threshold is 15$\sigma$ and no
explicit reobservations are made.  During the survey, a total of
$4\times 10^{13}$ channel-sky position observations have been made in
4500 hr of observations.

\section{BASIC ISSUES IN ANALYZING SURVEY EVENTS}\label{sec:basic}

In the following we compare alternative explanations for the candidate
events in a survey.  For definiteness, we cast the discussion in terms
of the META survey, though our formalism is general.  We stress the
fact that, in META, no signals were detected in any reobservations of
the relevant sky positions.

\subsection{Models for Survey Events}\label{sec:models}

The models we consider are:
\begin{description}
\item[{\bf Model~I:}] The measured signal consists only of radiometer noise
  in both the survey and the reobservations.
\item[{\bf Model~II:}] The measured signal results from noise combined with a
  real celestial signal of average strength $S$ that has scintillated
  into an improbably strong amplitude, with gain $g$ in the survey, but
  has remained below threshold in the reobservations.
  We assume that the signal would be constant were it not modulated by
  scintillations.
\item[{\bf Model~III:}] The survey detection results from 
  a real, {\em transient\/} signal
  of either celestial or terrestrial origin that does not repeat in any of
  the reobservations.\footnote{
   {One may compare the SETI case with gamma-ray bursts (GRB), which do
   not repeat, but have been identified as extraterrestrial sources.
   This identification relies on the fact that terrestrially produced
   GRBs that would interfere with the detection of astronomical bursts
   are rare, thankfully, and because many GRBs have been observed with
   more than one spacecraft, allowing time-of-arrival constraints on
   their directions.}}
  Terrestrial origins for the survey detection would
  include radio frequency interference (RFI) or cosmic ray events and
  failures in the electronics.    
  We characterize the transient signal in the survey as
  an amplitude $S_{\rm RFI}$.
\end{description}
In Fig.~\ref{fig:cls_ts} we show simulated time series for the three
models. 

Models~II and III are similar in that the survey is assumed to have 
acquired events that are extremely rare, whether due to atypical forms of RFI 
that pass through RFI-rejection filters or to 
rare scintillation peaks of real celestial sources.
Surveys that sift through tens of teraevents are indeed likely to 
yield such rare occurrences.

In later sections we shall compare, in a quantitative sense, the
viability of these three models in interpreting candidate events and
their evident nonrepeatability.  While it is tempting to dismiss the
events as a rare kind of RFI, we take the point of view that one
should be able to show quantitatively that Model~III is superior to
Model~II.  We also anticipate that future surveys will yield similar
events that will demand assessment and that some may eventually show
successful detections in the reobservations.  Therefore, we shall also
explore optimal methods for conducting and reporting future SETI
programs.

\subsection{General Considerations on Scintillation Amplitudes}
\label{sec:general_iss}

Some simple facts about scintillations and the demography of
transmitting sites in the Galaxy suggest that Model~II may not be 
dismissed so easily.  In the large number of trials conducted in META
($6\times 10^{13}$), it is remarkable that {\em only\/}  37 remaining
signals are unidentified and, for the most part, they satisfy some of
the requirements expected for a real ETI signal: a narrowband spectral
line in a rest frame of celestial interest.  If we interpret the
nonrepeatability of the candidates to be the result of scintillation
modulation, rather than to intrinsic variability, then it must be
assumed that the scintillation gain was extraordinarily high at the
time of the detection but was at more probable levels in all
reobservations.  In order for this to be a likely scenario, the number
of statistical trials must be large enough so as to make it probable
that a large scintillation gain occurred 37 times during the META
survey.  The probability of large gains, in turn, is closely related
to the number of signals transmitted from all ETI sites in the Galaxy.

As before, let $N_{\rm D}$ be the number of ETI civilizations in the
Galaxy, each transmitting $N_{\rm lines} \ge 1$
spectral lines.  These lines may or may not be in bands observed by
META or in any other survey.  A small multiplicity of lines is
advantageous for combatting interstellar scintillation
(\cite{cc95}; \cite{cs95}).  
If no ``magic'' frequencies or reference
frames are believed to pertain to the choice of frequency, the best we
can do {\em a priori\/}  is specify a (large) spectral domain of
bandwidth $B_{\rm D}$ for the possible frequencies of transmission.  
The total number of ``scintillation trials'' for a full sky survey
is
\begin{equation}
N_{\rm trials,ISS} \sim N_{\rm D} N_{\rm lines} K_{\rm s} P_{\nu},
\label{eq:trials_iss}  
\end{equation}
for $K_{\rm s}$ trials per sky position and where $P_{\nu}$ is the
probability that the observed frequency band in a survey does, in
fact, contain signals from ETI sources when a telescope is pointed
toward a given source.  In the optimistic case, where the choice of
frequency bands is believed to be the proper one and that one or more
of the transmitted lines for each and every civilization falls into
one of the chosen bands,
\begin{equation}
P_{\nu} = 1.
\label{eq:pnu_magic}
\end{equation}
However, if all frequencies in a large spectral domain 
$B_{\rm D}\gg\Delta\nu_{\rm spectrometer}$ are equally
probable, then
\begin{equation}
P_{\nu} \approx \frac{N_{\nu} N_{\rm frames} \Delta\nu_{\rm spectrometer}}
                     {B_{\rm D}} \ll 1.
\label{eq: pnu_equal}
\end{equation}

Let $P_g(g_{\rm max}) = \exp(-g_{\rm max})$ be the 
probability that a scintillation gain $g_{\rm max}$, or larger, occurs.  
Combined with the number of scintillation trials, we constrain
values of $g_{\rm max}$ using
\begin{equation}
N_{\rm trials, ISS} P_g(g_{\rm max}) \sim N_{\rm cand}
\label{eq:match}
\end{equation}
or
\begin{equation}
g_{\rm max} \sim \ln\biggl [\frac{N_{\rm D} N_{\rm lines} K_{\rm s} P_{\nu} }
                  {N_{\rm cand}} \biggr ].
\label{eq:gmax}
\end{equation}
Using values for META from the previous section
\begin{equation}
g_{\rm max} \sim \ln \left ( 0.11 N_{\rm D} N_{\rm lines} K_{\rm s}
     P_{\nu} \right ).
\label{eq:g_constraint}
\end{equation}
The meaning of Eq.~\ref{eq:g_constraint} is that if one wants to invoke
{\em likely\/} values for scintillation gains to account for
candidate events, the gains must not strongly  exceed a well-defined value,  
namely $g_{\rm max}$. 
Given that there must be 
at least as many transmitters as there are candidates, $N_{\rm D} \ge N_{\rm cand}$,
it may be concluded that  
$g_{\rm max}$ can range from a value of order unity if there are few signals
transmitted from sources in the Galaxy (in bands that have been observed)
to a value that is easily in the range of 10 to 20 if ETIs abound
in the Galaxy ($N_{\rm D} \gg 1$) or are profligate in their use of 
the radio spectrum ($N_{\rm lines} \gg 1$). 

If $g_{\rm max}$ is large, the intrinsic source strength can be
proportionately smaller.  For candidate events with intensities 
$I_{\rm cand}/\Nave \sim 32$--40, say, the intrinsic source strength
(i.e., in the absence of ISS)
need be only $I_{\rm cand}/(g_{\rm max}\Nave)$, which could be just a
few if $g_{\rm max} \sim 10$ to 20.  
Since the survey threshold in META was $\sim 32\Nave$ and the
reobservation threshold was $\sim 20\Nave$, the implication is that
common values of the scintillation gain will render the signal
undetectable, even with a large number of reobservations.  For
example, in 100 reobservations of a specific source, 
the maximum likely  scintillation gain is $\ln\,100
= 4.6$, yielding signal amplitudes 
$gS \lesssim g I_{\rm cand} / g_{\rm max} 
    \lesssim 15 \langle N \rangle$, 
too small to be detected in reobservations.

%

\subsection{General Considerations on Noise Amplitudes}
         \label{sec:general_noise}

Compared to the number of independent ISS tests, an enormous
number of independent trials is made with respect to noise. This number is
\begin{equation}
N_{\rm trials,noise}
 \sim N_{\rm ch} N_{\nu} N_{\rm frames} N_{\rm pol} K_{\rm s} N_{\rm sky},
\label{eq:trials_noise}
\end{equation}
where $N_{\rm sky}$ is the number of independent sky positions searched.
As stated before, $N_{\rm trials,noise}\sim 6\times 10^{13}$ for META.  For
the simplest spectral analysis of an unsmoothed FFT spectral estimate,
the noise in the spectrum is distributed exponentially
(c.f. Eq.~\ref{eq:pdf} with $S=0$) and the false-alarm probability is
specified so that $N_{\rm trials, noise} P_{\rm fa} \sim 1$.  For
META, the corresponding threshold is about $32\langle N\rangle$.  This
implies that much more modest amplitudes, say $16\Nave$, occur many
times
($\sim \sqrt{N_{\rm trials, noise}} \sim 10^7$ for META)
 during the survey.  We will find in our analysis below that
these modestly frequent noise spikes, combined with rare scintillation
gains, play a key role in determining the likelihood function for
signals from ETI sources.


\section{LIKELIHOOD FUNCTIONS OF META-LIKE SURVEYS}\label{sec:obs_events}

To quantify a survey, we define $N_{\rm sky}$ as the number of
independent sky positions (or telescope beam areas) considered;
$N_{\rm ch}$ the number of independent frequency channels; and 
$K_{\rm s}$ the number of independent observations or trials made per sky
position.  Consider a survey that yields $N_{\rm cand}$ candidate events
having amplitudes $I^{({\rm \survey})}(j)$ above a threshold 
$I^{({\rm \survey})}_{\rm T}$.  The results of the survey may be described as
the set of candidate signal intensities

\begin{equation}
{\cal C} = 
 \left\{ I^{({\rm \survey})}(j) \ge \threshsu, 
      j = 1, \ldots, N_{\rm cand}\, \right\}
\label{eq:survey} 
\end{equation}
and the non-detections
\begin{equation}
{\cal N} = \left \{
           I^{(\survey)}(j) < \threshsu, j = 1, \ldots, \Nsurvey - N_{\rm cand} 
           \right \},
\label{eq:survey_non}
\end{equation}
where $\Nsurvey \equiv N_{\rm trials,noise}$. 
Further,  assume that the direction for each candidate survey event 
is reobserved
$K_{\rm r}$ times and that no signal is found above a threshold intensity
$I^{({\reobs})}_{\rm T}$, yielding the set of non-detections during 
reobservations,
\begin{equation}
{\cal R} 
 = \left \{ {K_{\rm r}}(j)\times N_{\rm ch}^{(\reobs)} \,\,{\rm amplitudes} \,\, < \threshro, 
     j=1, \ldots, N_{\rm cand} \right \}.
\label{eq:reobs_non}
\end{equation}
Because the total number of statistical tests in reobservations
is much smaller than in the survey, the reobservation threshold may
be significantly smaller than in the survey.
For simplicity in our discussion, we
assume the threshold of all reobservations to be the same, but this need not be
the case.  It is a simple matter to account for thresholds that vary due
to use of different dwell times or different background noise levels.

We distinguish reobservations where the scintillation gain is correlated
with the gain, $g_{\rm c}$, during the original candidate detection from those
where it was uncorrelated.  In practice, without explicit detection of
an ETI signal over a sustained time, one does not know how long
the gain is correlated because it depends on the strength of scattering
to the particular source.   Nonetheless, statistical inference from 
a set of observations (viz. an initial detection with only upper bounds
in reobservations, as in META) depends on the degree of correlation
assumed.  We will make the simplifying assumption that
reobservations have either {\em completely\/} correlated
or completely uncorrelated gains.  Consequently, we write
the number of reobservations as the sum of correlated and uncorrelated
numbers,
\begin{equation}
{K_{\rm r}}(j) = \krc(j) + \kru(j).
\label{eq:ro_split}
\end{equation}


The likelihood function for a given, {\em individual\/} 
candidate event is
\begin{equation}
{\cal L} = {\cal L}^{({\rm \survey})} {\cal L}^{({\rm \reobs})},
\label{eq:like}
\end{equation}
where ${\cal L}^{({\rm \survey})}$ is proportional to
 the probability density evaluated at the 
event's survey amplitude 
$I^{({\rm \survey})}$ and ${\cal L}^{({\rm \reobs})}$ 
is the probability that no detection was made in reobservations.   For the 
three models presented in \S\ref{sec:models}, the survey likelihoods are
(dropping the $j$ subscript for clarity)
\begin{eqnarray}
\cl^{(\survey)}_{\rm I} &\approx& 
          \Delta I f_{I}\left(I^{(\survey)}; S=0\right) \nonumber \\
\cl^{(\survey)}_{\rm II} &\approx& 
          \Delta I f_{I}\left(I^{(\survey)}; g_{\rm c}S\right) \\ 
\cl^{(\survey)}_{\rm III} &\approx& 
          \Delta I f_{I}\left(I^{(\survey)}; S_{\rm RFI}\right)
            \nonumber 
\label{eq:like_factors_s},
\end{eqnarray}
where $f_{\rm I}$ is defined in \S\ref{sec:pdfs} and
Appendix \ref{app:stats};
$\Delta I \ll I^{(\survey)}$ is the uncertainty in establishing the
survey amplitudes and $g_{\rm c}$ is the scintillation gain
{\em at the time of the candidate detection}.  
The approximate equalities in Eq.~\ref{eq:like_factors_s}
result because the exact form would be integrals over intensity
of $f_{\rm I}$
over intervals $\Delta I$ centered on the candidate intensity.
The reobservation likelihoods are the probabilities of no
detections in $K_{\rm r}$ reobservation trials:
\begin{eqnarray}
\cl^{({\rm \reobs})}_{\rm I} &=& 
 \left [ 1 - P_{\rm d}\left(I^{({\rm \reobs})}_{\rm T}, 0\right) \right ]^{K_{\rm r}} 
 \nonumber \\ 
\cl^{(\creobs)}_{\rm II} &=& 
 \left [ 1 - P_{\rm d}\left(I^{(\reobs)}_{\rm T}, g_{\rm c}S\right) \right ]^{{\krc}} \\ 
\cl^{(\ureobs)}_{\rm II} &=& 
 \left [ 1 - P_{\rm d,scint}\left(I^{(\reobs)}_{\rm T}, S\right) \right ]^{{\kru}} \\ 
\cl^{(\reobs)}_{\rm III} &=& 
   \left [ 1 - P_{\rm d}\left(I^{(\reobs)}_{\rm T}, 0\right) \right ]^{K_{\rm r}}. 
   \nonumber 
\label{eq:like_factors_r}
\end{eqnarray}
Detection probabilities were introduced in \S\ref{sec:pdfs} and
Appendix \ref{app:stats}.

For Model~II, we have written separate, reobservation likelihood functions
for correlated and uncorrelated scintillation gains. 
It is useful to normalize intensities, thresholds, and signal
strengths by the mean noise level $\Nave$, e.g.,
\begin{eqnarray}
\eta &=& I / \Nave \nonumber \\
\zeta &=& S / \Nave.
\label{eq:normalization}
\end{eqnarray}
The likelihood functions become 
\begin{eqnarray}
\cl_{\rm I} &=& 
  \cl^{(\survey)}_{\rm I} \cl^{(\reobs)}_{\rm I} = 
  {\rm e}^{-\eta} 
   \left[1 - {\rm e}^{-\eta^{(\reobs)}_{\rm T}}\right]^{K_{\rm r}}\nonumber \\
\cl_{\rm II} &=& 
   \cl^{(\survey)}_{\rm II} \cl^{(\reobs)}_{\rm II} =
     I_0\left(2\sqrt{\eta g\zeta}\right) 
   {\rm e}^{-(\eta + g\zeta)} 
   \left[1- P_{\rm d}\left(\eta_{\rm T}^{(\reobs)}, g_{\rm c}S\right) \right]^{\krc}
   \left[1-{\rm e}^{-\eta^{(\reobs)}_{\rm
T}/(\zeta+1)}\right]^{{\kru}} \nonumber \\
\label{eq:like_models}
\cl_{\rm III} &=& 
   \cl^{(\survey)}_{\rm III} \cl^{(\reobs)}_{\rm III} = 
    I_0\left(2\sqrt{ \eta \zeta_{\rm RFI} }\right) 
    {\rm e}^{-(\eta + \zeta_{\rm RFI})}
   \left[1 - {\rm e}^{-\eta^{(\reobs)}_{\rm T}} \right]^{K_{\rm r}}, 
\end{eqnarray}
where $I_0$ is the Bessel function and we have dropped a 
(constant) factor 
$(\Delta I / \Nave)$ on the right-hand side of each equation.


\section{INDIVIDUAL CANDIDATE EVENTS WITH NO CORRELATED
  REOBSERVATIONS}\label{sec:like_events_u}

In the surveys summarized earlier, 
unexplained, extrastatistical events are found in a few  directions
that, when reobserved, yield no evidence for  persistent signals.

Here, we consider multiple reobservations of a given sky
position (i.e., an {\it individual} candidate source)
that are completely statistically independent.  For the case
of scintillating sources, this means that reobservations are made more
than one scintillation time after an initial detection.  While the
META survey did, in fact, make ``prompt'' reobservations only 40 s
after initial candidates were found, it is useful,for
pedagogical reasons and for purposes of generality, to analyze events
as if they were statistically independent.  It is possible, for
example, that the reobservations are, in fact, statistically
independent if ETI sources are rare and are at large distances, thus
producing short scintillation times (i.e., $\lesssim 40$ s).  Moreover,
we show in \S\ref{sec:like_events_c} that correlated reobservations
combined with uncorrelated ones do not change substantially the
conclusions we find here.  In \S\ref{sec:like_survey} we apply a
likelihood analysis to an entire survey, which allows constraints to
be made on the overall population of ETI transmitters that may 
exist in the Galaxy.

\subsection{Comparison of Likelihood Functions for Individual Events}
   \label{sec:comments}

Some initial conclusions may be made from direct inspection of the
likelihood functions.  Firstly, given that survey events
necessarily have large amplitudes 
(above the false-alarm threshold and, thus, 
well above the mean noise level $\Nave$), the noise-only Model~I has
a low likelihood compared to the other two models.
This becomes clear for events $\eta \ge \eta_{\rm T} \gg 1$
when we approximate $I_0(x) \sim e^x / \sqrt{2\pi x}$
and calculate the ratio (dropping the `RFI' subscript on $\zeta$):
\begin{equation}
\frac {\cl_{\rm III}} {\cl_{\rm I}} \sim
     \frac{\exp (2\sqrt{\eta\zeta}-\zeta)}
          {(4\pi\sqrt{\eta\zeta})^{1/2}}.
\label{eq:ratio31}
\end{equation}
For signal strengths $1 < \zeta \lesssim \eta$ and 
candidate strengths $\eta \ge 32$, we have
$\cl_{\rm III} / \cl_{\rm II} > 10^3$.
Similarly, Model II can be shown to be far more likely than Model I for
many parameter values. 
In general, 
for $\eta \sim \zeta_{\rm RFI} \sim g_{\rm c}\zeta \gg 1$ and $g_{\rm c} \gg 1$
we find that Models~II and III are much more strongly preferred than
Model~I.    For sufficiently small scintillation gains, $g_{\rm c}$, however, 
$\zeta$ can be large enough that the reobservation
factor $\cl_{\rm II}^{(\rm ro)} / \cl_{\rm I}^{(\rm ro)}$ will make Model~II 
less likely than even Model~I.

Secondly, if we hypothesize, in Models~II and III, that a strong
source (real or RFI) underlies the event and that, whatever its
nature, the amplitude was the same in either model (i.e., 
$g_{\rm c}\zeta = \zeta_{\rm RFI}$), then $\cl^{({\survey})}_{\rm II} \equiv
\cl^{({\survey})}_{\rm III}$, but the net likelihood ratio is
(assuming only uncorrelated reobservations for model II)
\begin{equation}
r \equiv\frac{\cl_{\rm II}}{\cl_{\rm III}} 
 = \frac{\cl^{(\reobs)}_{\rm II}}{\cl^{(\reobs)}_{\rm III}}
 = \left [ \frac {1 - {\rm e}^{-\eta^{(\reobs)}_{\rm T} / 
  (\zeta + 1)}} {1 - {\rm e}^{-\eta^{(\reobs)}_{\rm T}}} \right ]^{\kru} 
 \le 1.
\label{eq:ratio23}
\end{equation} 
The larger any real signal of average, continuous strength $\zeta$ is
in Model~II, the smaller the likelihood ratio in Eq.~\ref{eq:ratio23}
and the more that Model~II is disfavored.  However, if the survey
detection was made when $g_{\rm c}$ was far out on the tail of the
exponential pdf, then $\zeta$ can be sufficiently small that 
$\cl_{\rm II} / \cl_{\rm III}\to 1$.

For a specified lower limit on $r$, e.g., $r \ge 0.5$, one obtains a
likely upper bound on $\zeta$
\begin{equation}
\zeta \le \zeta_{\rm max} \equiv 
     \frac 
        {\eta_{\rm T}^{(\reobs)} }
        {\biggr\vert 
          \ln\left[1-r^{1/\kru} \left(1-{\rm e}^{-\eta_{\rm T}^{(\reobs)}}\right)\right]
         \biggr\vert } - 1.  
\label{eq:zeta_range}
\end{equation}
For $\kru = 256$ (typical of META) and $r \ge 0.5$, we obtain $\zeta
\le 2.38$ using a reobservation threshold of $\eta_{\rm T}^{(\reobs)} =
20$.  For a typical candidate signal strength, $\eta = 32$, a
scintillation gain $g_{\rm c} = \eta / \zeta \ge 13.4$ 
is required in the survey.
The probability of this occurring is small, 
$P_g(13.4) = {\rm e}^{-13.4} = 10^{-5.8}$, 
but it is large enough that in a many-trial
survey, such a gain is likely to be encountered  
if $N_{\rm trials, \,\, ISS}$ is large, as discussed in
\S\ref{sec:general_noise}.

Therefore, with measurements such as those outlined in
\S\ref{sec:obs_events}, it is impossible to favor Model~II over
Model~III.  While $P_g(13.4) = 10^{-5.8}$ is not large, given that we
do not know the number of transmitters in the Galaxy and what $g_{\rm
max}$ might have been in the survey, we also cannot favor Model~III
over Model~II.  Statistically, over a set of events, however, it may
be possible to conclude that an unlikely ensemble of scintillation
gains is required to account for the known events.  Also, the likelihood
does not automatically account for the probability that particular
scintillation gains will occur.  For this reason, we extend our
analysis in the next section to include Bayesian factors.

\subsection{Bayesian Analysis of Individual Events}\label{sec:bayes_events}

In this section we adopt a Bayesian analysis for the {\em a
posteriori\/} joint pdf of $g_{\rm c}$ and $\zeta$ for a particular
candidate
event.  The Bayesian analysis proceeds by deriving {\em a posteriori\/}  
pdf's for the unknown parameters $g_{\rm c}, \zeta, \zeta_{\rm RFI}$
by taking the product of the appropriate likelihood function (for a
given model) with {\em prior\/}  pdfs for the parameters and normalizing
by the integral of the product over the parameter space.
The advantage of this analysis is that it penalizes improbably 
large scintillation gains and large source fluxes.

The {\em a posteriori\/}  pdf for a vector of parameters
\mbox{\boldmath{$\theta$}} is, in terms of the survey detection with
amplitude $\eta_{\rm c}$ and nondetections in reobservations,
\begin{equation}
f(\mbox{\boldmath $\theta$} \vert \eta_{\rm c} {\cal I})
 = \frac{ f_{\mbox{\boldmath $\theta$}}(\mbox{\boldmath $\theta$}) \cl (\eta_{\rm c}\vert
\mbox{\boldmath $\theta$} {\cal I}) }
     {f(\eta_{\rm c} \vert {\cal I}) },
\label{eq:b1}
\end{equation}
where $\cl$ represents one of the likelihood functions defined
previously\footnote{
Following standard practice, we explicitly show the background
information as an element of the analysis.  However, nonstandardly, we
use ``pdf'' notation for the {\em posterior\/} and {\em prior\/}  pdfs
for the parameters rather than using the common, but misleading, $p$
notation.  We also write the likelihood function with arguments that
often are written only when the identical quantity is referred to as
the sampling distribution (e.g., \cite{gl92}).
};
$f_{\mbox{\boldmath $\theta$}}(\mbox{\boldmath $\theta$})$ is the joint {\em a priori\/}  pdf for the
parameters \mbox{\boldmath{$\theta$}}; ${\cal I}$ represents background information; and
the denominator 
\begin{equation}
f(\eta_{\rm c}\vert{\cal I}) \equiv \int \, d\mbox{\boldmath $\theta$}
\,f_{\mbox{\boldmath $\theta$}}(\mbox{\boldmath $\theta$})
    \cl(\eta_{\rm c}\vert \mbox{\boldmath $\theta$} {\cal I})
\label{eq:norm1}
\end{equation}
normalizes the pdf.

Model~I for candidate events (noise only) has no parameters other
than the background noise level, which we assume to be known.  

In Model~II, the two parameters to be determined for each event are
the scintillation gain, $g_{\rm c}$, and the intrinsic (average)
signal strength, $\zeta$.  We assume $g_{\rm c}$ and $\zeta$ are
statistically independent in the joint {\em prior\/}  pdf, which therefore
factors into the individual pdf's.  The {\em prior\/}  pdf for $g_{\rm c}$ is
$f_g(g_{\rm c}) = \exp(-g_{\rm c})$, appropriate for the strong
scattering regime, which we assume.
For $\zeta$ (i.e., $S$) we adopt a pdf that corresponds to the
power-law distribution of sources considered in \S\ref{sec:candles}
(cf. Eq.~\ref{eq:pdf_candles}).  We assume a Galactic disk population
of sources ($\alpha = 2$) and take the range of source strengths to be
large, $\zeta_1 \gg \zeta_0$ (or $S_1 \gg S_0$).  The {\em prior\/}  pdf
for $\zeta$ is then
\begin{equation}
f_{\zeta}(\zeta)  \approx
\zeta_0\zeta^{-2}\,U(\zeta-\zeta_0)U(\zeta_1-\zeta)
\label{eq:fzeta2}
\end{equation}
where we have used Heaviside functions $U$ to define the cutoffs of
the power-law pdf.

The {\em posterior\/}  pdf for Model~II then becomes
\begin{equation}
f_{\rm II}(g_{\rm c},\zeta \vert \eta_{\rm c} {\cal I}) =  \frac
   {f_g(g_{\rm c}) f_{\zeta}(\zeta) \cl_{\rm II} (\eta_{\rm c} \vert g\zeta {\cal I}) }
   {f(\eta_{\rm c} \vert {\cal I} )}.
\label{eq:postIIa}
\end{equation}
For the assumed {\em prior\/}  pdfs (and dropping the Heaviside functions for
clarity), we have
\begin{equation}
f_{\rm II}(g_{\rm c},\zeta \vert \eta_{\rm c} {\cal I}) 
\propto 
{\rm e}^{-(g_{\rm c}+g_{\rm c}\zeta+\eta_{\rm c})} 
    \zeta^{-2} I_0\left(2\sqrt{g_{\rm c}\zeta\eta_{\rm c}}\right)
    \left[1 - {\rm e}^{-\eta_{\rm T}^{(\reobs)}/(1+\zeta) } \right]^{\kru},
\label{eq:postIIb}
\end{equation}
where the proportionality constant is $\zeta_0/{f(\eta_{\rm
c}\vert{\cal I})}$.  Strictly, the cutoffs $\zeta_{0,1}$ of the source
strength distribution are also parameters to be estimated or
constrained and they could be included explicitly in the {\em
posterior\/} pdf, necessitating inclusion of corresponding {\em prior\/} 
pdfs on the right hand side of Eq.~\ref{eq:postIIa}--\ref{eq:postIIb}.
For now, however, we treat these as hidden parameters, assuming that
the range of source strengths, $\zeta_0 \le \zeta \le \zeta_1$, is
large enough to include the best fit values of $g_{\rm c}$ and
$\zeta$.

For Model~III, which considers candidate events in the survey
to arise from events that are intrinsically transient, we hypothesize
a distribution of RFI amplitudes, $f_{\rm RFI}(\zeta_{\rm RFI})$,
that has parameters to be determined via application of Bayes' theorem.
As an illustration, we consider the simple example of a 
pdf that is constant in the interval 
$\zeta_{0, {\rm RFI}} = 0 \le \zeta_{\rm RFI} \le \zeta_{1,{\rm RFI}}$,
\begin{equation}
f_{\rm RFI}(\zeta_{\rm RFI})
 = \frac{1}{\zeta_{1,{\rm RFI}}}U(\zeta_{\rm RFI})U(\zeta_{1,\rm RFI}-\zeta_{\rm RFI}).
\label{eq:priorRFI}
\end{equation}
With the exception of multiplying  constants, the resulting {\em posteriori\/}  
distribution does not differ from ${\cal L}_{\rm III}$.

\subsubsection{Application to META}\label{sec:meta1}

Of the candidates reported by HS93, the majority
of the extrastatistical events are near $\eta_{\rm c} = 32$ with a few
at larger $\eta_{\rm c}$. 
In Fig.~\ref{fig:cls_eventpdf} we show
contours of the {\em posterior\/}  pdf, Eq.~\ref{eq:postIIb}, for four
cases:  $\eta_{\rm c} = 32$ and 100 with thresholds $\eta_{\rm
T}^{(\reobs)} = 10$ and 20; in all cases we assume that $\kru = 256$
reobservations were attempted.


Table~2
displays the values of $g_{\rm c}$ and
$\zeta$ that maximize the {\em posterior\/}  pdf for the
specific $\eta_{\rm c}$ and $\eta_{\rm T}^{(\reobs)}$ we have
considered here.  In general, we find that as $\eta_{\rm c}$
increases (for constant $\eta_{\rm T}^{(\reobs)}$), the peak value of
the pdf decreases, the region of constant probability becomes more
concentrated about the peak, and the peak moves to larger gain and
slightly larger signal strength.

A significant outcome of this analysis is that, at the time of the
initial detection, the {\em apparent source strength}, $g_{\rm
c}\zeta$, is only about 50\% of the measured candidate strength.  The
implication is that, even at these large intensities, the noise
contributes significantly and sources with intrinsic intensities well
below threshold will most likely be seen as the {\em combination\/}  of
a large scintillation gain and a large noise spike.  Although large,
the scintillation gains derived here are considerably less than those
derived in our earlier discussion of the likelihood functions for
Models~II and III (cf. \S\ref{sec:comments}) where an observed
intensity of $\eta_{\rm c} = 32$ implied a scintillation gain $g_{\rm
c} \gtrsim 13.4$.  These smaller scintillation gains are considerably
more probable, $P(g \ge 6.2) = {\rm e}^{-6.2} = 10^{-2.7}$.


Figure~\ref{fig:cls_gz_rho0} shows the best fit values for $g_{\rm
c}$, $\zeta$, and $g_{\rm c}\zeta$ as a function of the number of uncorrelated
reobservations, $\kru$.  We show results for three values of the
reobservation threshold, $\eta_{\rm T}^{(\reobs)} = 5$, 15, and 20.
In all cases, the product $g_{\rm c}\zeta$ is less than the
apparent source strength at the time of the detection.  This figure
demonstrates two effects: (1) that noise {\em in general\/}  plays a key
role in producing candidate events in a large scale survey; and (2)
even a large number of reobservations fails to constrain the source
flux to be very small unless the reobservation threshold $\eta_{\rm
T}^{(\reobs)} \sim 5$, much less than that used in META.

Our conclusions may be understood by referring to the various factors
in Eq.~\ref{eq:postIIb}.  For event amplitudes ($\eta_{\rm c}$) of interest,
the argument of the Bessel function is large,
$2\sqrt{g_{\rm c}\zeta\eta_{\rm c}} \gg 1$.  Using the large-argument approximation
for $I_0$, we may write the {\em posterior\/}  pdf for $g_{\rm c}$ and $\zeta$ as
\begin{equation}
f_{\rm II}(g_{\rm c},\zeta \vert \eta {\cal I}) \propto 
\left[{\rm e}^{-g_{\rm c}}\right ]
\left[ \zeta^{-2} \right ]
\left[ {\rm e}^{-\left(\sqrt{\eta_{\rm c}} - \sqrt{g_{\rm c}\zeta}\right)^2} 
             \left (\eta_{\rm c} g_{\rm c}\zeta\right)^{-1/4} \right ]
\left[1 - {\rm e}^{-\eta_{\rm T}^{(\reobs)}/(1+\zeta) } \right]^{\kru}.
\label{eq:postIIc}
\end{equation}
The third and fourth bracketed factors in Eq.~\ref{eq:postIIc}
represent the likelihood function $\cl_{\rm II}$ and, by themselves,
are maximized for $g_{\rm c}\zeta \sim \eta_{\rm c}$ and $\zeta\ll 1$,
respectively.  Maximizing the {\em likelihood function\/}  therefore
yields an apparent source flux $g_{\rm c}\zeta$ that is nearly equal
to the observed candidate event $\eta_{\rm c}$ but favors actual
source strengths $\zeta$ that are small; hence, the ML value for
$g_{\rm c}$ is large.  The ML estimate for $g_{\rm c}$ is unrealistic
because it is unlikely to have occurred in the number of trials
considered, unless ETI is hyperabundant in the Galaxy.  The Bayesian
estimation procedure is far more realistic because it builds in
penalties for both large scintillation gains and strong sources via
the first two factors in Eq.~\ref{eq:postIIc}.  These factors favor,
respectively, small $g_{\rm c}$ and small $\zeta$.  We then see that
it is through these two factors that the Bayesian estimates yield
apparent source fluxes that are considerably smaller than the measured
event amplitude.

The {\em a posteriori\/} pdf also displays, for small $\eta_{\rm
T}^{(\reobs)}$, an important bimodality, as seen in
Fig.~\ref{fig:cls_eventpdf}{\it c}.  One mode occurs at ``large''
$g_{\rm c}$ and $\zeta$, $g_{\rm c} > 1$ and $\zeta > 1$, the other at
the origin of the contour plot.  The bimodality occurs once the number
of reobservations $\kru$ becomes large enough that the fourth factor
in Eq.~\ref{eq:postIIc} compensates the third factor.  For still
larger $\kru$, the reobservation factor dominates and only the maximum
near the plot origin remains.  This ``collapse'' of the posterior pdf
is a consequence of the particular source strength distribution,
$f_S(S)$, we have adopted, namely a disk population of
standard candles, which favors $\zeta \to 0$ once the fourth factor
dominates the third factor.  In practice, we search for the maximum of
$f_{\rm II}(g_{\rm c},\zeta \vert \eta {\cal I})$ with a minimum
$\zeta > 0$; for Fig.~\ref{fig:cls_eventpdf}, we have searched $\zeta
\ge 0.1$.


\subsection{A Critical Number of Reobservations}\label{sec:bayes_critical}

The bimodal form of the {\em posterior\/} pdf in Eq.~\ref{eq:postIIb}
may be used to derive the number of reobservations needed to rule out
Model~II, where the source strength, $\zeta$, is assumed constant in
time.  For META parameters, we have seen in the previous section that
a small number of reobservations influences the parameter estimates
for $g_{\rm c}$ and $\zeta$ but they do not rule out Model~II.  Since,
in the limit $\zeta\to 0$, Model~II becomes identical to Model~I, our
approach will be to estimate the number of reobservations, $\kru$,
required to make $\zeta \ll 1$ the most probably value for $\zeta$.

The logarithm of the {\em posterior\/}  pdf, Eq.~\ref{eq:postIIb}, is
\begin{equation}
\lambda_{\rm II}
 = -g_{\rm c} - 2\ln\zeta + \ln f_I(\eta_{\rm c}, g_{\rm c}\zeta) 
 + \kru\ln \left [1 - P_{\rm d,scint}\left(\eta_{\rm T}^{(\reobs)}, \zeta\right) \right ].
\label{eq:logbayes}
\end{equation}
For the case where there
are no reobservations ($\kru = 0$), $\lambda_{\rm II}$ is maximized
for $g_{\rm c} = 2$ and $\zeta = \eta_{\rm c}/2 - 9/4$.  The
difference between $\max(\lambda_{\rm II})$ and its value for small
signal strength, $\zeta \lesssim 1$, is $\sim \eta_{\rm c}$.  For
reobservations to affect the location of the maximum significantly,
the last term in Eq.~\ref{eq:logbayes}, which favors small $\zeta$,
must alter the slope $\partial\lambda_{\rm II}/\partial\zeta$ so that
it becomes negative for all $\zeta \gtrsim 1$.  For such negative
slopes, there cannot be a ``large'' $\zeta$ maximum in $\lambda_{\rm II}$ 
that would imply a real signal was present.  By maximizing
$\lambda_{\rm II}$ with respect to $g_{\rm c}$ and requiring a
negative slope at $\zeta=1$, we obtain the number of reobservations
needed to rule out a detectable signal.  The result is $\kru \ge
\krucrit$ where the critical number of reservations is, approximately,
for $\eta_{\rm c} \gg 1$,
\begin{equation}
\krucrit \approx  
 \left[\frac{\partial}{\partial\zeta} \ln (1 - P_{\rm d,\rm scint}) \right ]^{-1}
 \left(\frac{\eta_{\rm c} - 9}{4} \right )
\approx 
\frac{ {\rm e}^{\eta_{\rm T}^{(\reobs)}/2} }  {\eta_{\rm T}^{(\reobs)} } 
    \left(1 - {\rm e}^{-\eta_{\rm T}^{(\reobs)}/2} \right )
    (\eta_{\rm c} - 9).
\label{eq:krucrit}
\end{equation}

In Fig.~\ref{fig:cls_krucrit} we show $\krucrit$ plotted against the
reobservation threshold $\eta_{\rm T}^{(\reobs)}$ for several values of 
candidate strength $\eta_{\rm c}$.  The figure demonstrates that
large thresholds produce exponentially large numbers of
reobservations required  to rule out Model~II.    
For META, where a reobservation
threshold $\eta_{\rm T}^{(\reobs)}\approx 20$ was used,
$\krucrit \approx 10^{4.3}$ for candidate events just above the
survey threshold, $\eta_{\rm T}^{(\survey)} = 31.7$.  Stronger candidate events
require a greater number of reobservations.  

{\em 
We conclude that candidate events in META cannot be ruled out as
being due to scintillating sources, even if they transmit with
constant power aimed into our direction, because there were
too few reobservations and the threshold was too high.
}

We also conclude that constraining reobservations must be made at
lower thresholds than have been used (cf. Fig.~\ref{fig:cls_p2cond}).
However, a lower reobservation threshold yields a larger number of
false alarms.  The false-alarm probability in reobservations is
$\exp\left[-\eta_{\rm T}^{(\reobs)}\right]$.
The number of false alarms expected in the critical number of
reobservations (per frequency channel) is
\begin{equation}
\Delta N_{\rm fa} = \krucrit P_{\rm fa}^{(\reobs)}
 = \frac{{\rm e}^{-\eta_{\rm T}^{(\reobs)}/2} }{\eta_{\rm T}^{(\reobs)}}
    \left(1 - {\rm e}^{\eta_{\rm T}^{(\reobs)}} \right) (\eta_{\rm c}-9).
\end{equation}
Figure~\ref{fig:cls_krucrit} also shows $\Delta N_{\rm fa}$ plotted
against reobservation threshold.  The dashed line shows where one
false alarm is expected per frequency channel.  For $\eta_{\rm c} =
32$ and $\eta_{\rm T}^{\rm (\reobs)} = 5$, $\Delta N_{\rm fa} \approx
0.38$.  While this number is small on a {\em per-channel\/} basis,
when combined with the fact that reobservations need to search in
frequency space to take into account unknown doppler shifts, it
suggests that many false alarms are to be expected in reobservations
made at small, otherwise constraining thresholds.

\section{INDIVIDUAL CANDIDATE EVENTS WITH CORRELATED
	REOBSERVATIONS}\label{sec:like_events_c}

In \S\ref{sec:bayes_events} the Bayesian estimates for scintillation
gain $g_{\rm c}$ and source strength $\zeta$ for individual events
were made while assuming that the scintillation gain had decorrelated
completely at the reobservation times.  This corresponds to
reobservations that are made hours or perhaps even just minutes after
an initial detection.  In META, however, a few reobservations were
made within 40~s of the initial detection, a time short enough that
the scintillations could have been highly, if not perfectly,
correlated between the first few reobservations and candidate
detection.  In this section we consider such prompt, correlated
reobservations in the Bayesian estimate of signal parameters in the
context of Model~II.

We assume, as before, that a candidate signal with amplitude,
$\eta_{\rm c}$, is reobserved in $\kru$ unsuccessful trials
where the scintillations are uncorrelated.  In addition, we include
$\krc$ reobservations, also unsuccessful, where the
scintillations are correlated over the time spanning the
reobservations and the initial detection.  To illustrate the effects
of such correlated reobservations, we assume they are 100\%
correlated.  In reality, the degree of correlation would vary slowly
from 100\% to zero correlation.  But the point here is to demonstrate
the stronger influence of perfectly correlated reobservations.  
We find, in
fact, that the Bayesian estimates for $g_{\rm c}$ and $\zeta$ are
quantitatively different when correlated observations are included.
However, the same conclusion is reached as before: the candidate
events occur (if Model~II is assumed) because noise and signal
conspire to produce a threshold crossing in the survey.  Moreover,
though the number of uncorrelated reobservations needed to rule out
Model~II decreases when one or more correlated reobservations is made,
the required number is still rather large unless smaller-than-used
thresholds are used in the reobservations.
 
Unsuccessful, correlated observations introduce a new factor in
the likelihood function and {\em a posteriori\/} pdf 
(cf. Eq.~\ref{eq:postIIa}--\ref{eq:postIIb}),
$\left [ 1 - P_{\rm d}(\threshro, g_{\rm c} \zeta)\right ]^{\krc}$
or, equivalently,  a term in the log pdf, 
\begin{equation}
\lambda_{\rm II}^{(\creobs)}
 = \krc \ln\left [1 - P_{\rm d}\left(\threshro, g_{\rm c}\zeta\right) \right ].
\label{eq:c_factor}
\end{equation}
This term favors small values of $g_{\rm c}\zeta$.  When included with
the other factors in the {\em posterior\/} pdf, it pushes estimates
for $g_{\rm c}\zeta$ to values that are {\em lower\/} than if there
were no correlated observations.  The influence of correlated
reobservations is illustrated in Fig.~\ref{fig:cls_gz_rho100}
(cf. Fig.~\ref{fig:cls_gz_rho0}), where estimates for $g_{\rm c}$,
$\zeta$, and $g_{\rm c}\zeta$ are plotted against $\kru$, for several
values of $\krc$, with thresholds $\eta_{\rm T}^{(\reobs)} = 20$, 10,
and 5.  The effects of the $\lambda_{\rm II}^{(\creobs)}$ term are to
decrease the best fit $\zeta$ as $\krc$ increases.  Also, the critical
number of uncorrelated reobservations needed to favor $\zeta \ll 1$
decreases as $\krc$ increases.  For the larger reobservation
thresholds, $\eta_{\rm T}^{(\reobs)} = 20$ and 10, the necessary
number of reobservations still measures in the hundreds, even for
$\krc = 16$ correlated reobservations.  With the smaller threshold,
only a few (but more than one) correlated reobservations cause the
best fit $\zeta$ to collapse below unity.

In Fig.~\ref{fig:cls_krccrit}, analogous to Fig.~\ref{fig:cls_krucrit},
we show the number of correlated reobservations needed to rule out Model~II.
Not surprisingly, this number is substantially smaller than the number of
correlated reobservations.    In practice, of course, the scintillation
time scale limits how many correlated observations may be done.  
A general expression for the number of uncorrelated and correlated
reobservations needed to rule out Model~II is
\begin{equation}
\frac{\partial}{\partial\zeta} 
   \left [
          P_{\rm d,scint}^{\kru} + P_{\rm d}^{\krc}
   \right ]_{\zeta=1, g\sim \eta_{\rm c}/4}
 > \left ( \frac{\eta_{\rm c} - 9}{4} \right ).
\label{eq:cucrit}
\end{equation}

We conclude that:
\begin{enumerate}
\item correlated but unsuccessful
reobservations reduce estimates for $g_{\rm c}$ and $\zeta$ from those seen
in Fig.~\ref{fig:cls_gz_rho0} where only uncorrelated reobservations
were assumed;  
\item  the occurrence of unsuccessful, correlated reobservations 
does not obviate
the conclusion that initial survey detections result from a 
noise-scintillation conspiracy where both noise and signal are needed
to provide the survey detection; and
\item  the noise-scintillation conspiracy makes it unlikely to 
detect the signal in a small number of reobservations. 
\end{enumerate}

\section{GLOBAL ANALYSIS ON ENTIRE SURVEYS} \label{sec:surveys}

In this section we return to the question we raised in
\S\ref{sec:comments}:  Although radiometer noise alone
cannot explain the extrastatistical events seen in META and similar
surveys, how likely is the scintillation interpretation?  Restated, is
the required {\em ensemble\/} of scintillation gains consistent with the gain
pdf {\em and\/} with a plausible distribution of intrinsic source
strengths?


\subsection{Survey Intensity Distribution}\label{sec:survey_pdf}

As in \S\ref{sec:like_events_u}, we take $N_{\rm cand}$ to be the number of
candidate signals, i.e., those intensities above threshold, in the
survey with the signal intensities in the set $\cal C$ (cf.
Eq.~\ref{eq:survey}).  These intensities are archived while the
remaining $\Nsurvey - N_{\rm cand}\ll N_{\rm cand}$ 
observations are discarded, so that the
only information for these observations is that the intensity was less
than $\threshro$.

Now suppose that some fraction $\epsilon_{\rm sky}$ of the beam areas
covered in the survey contain transmitting civilizations whose
received signal strengths $S$ ({\em before\/} accounting for
scintillation modulations) are described by a pdf, $f_S(S)$, and that
some fraction $\epsilon_{\rm ch}$ of the frequency channels surveyed
contain power from broadcasting civilizations.  
In terms of previously defined quantities, 
we may write
\begin{equation}
\epsilon_{\rm sky} = 1 - \langle {\rm e}^{-N_{\rm b}} \rangle,
\label{eq:esky}
\end{equation}
where $N_{\rm b}$ is the average number of sources expected per beam
(c.f. Eq.~\ref{eq:n_in_beam_mfp}) and the exponential is averaged
over all directions. 
The spectral fraction is
\begin{equation}
\epsilon_{\rm ch} = 
    \frac{N_{\rm signals/beam} \Delta\nu_{\rm spectrometer}} {B_{\rm D} N_{\rm ch}}
    \lesssim \frac{N_{\rm signals/beam}} {N_{\rm ch}},
\label{eq:ech}
\end{equation}
where $N_{\rm signals/beam} = N_{\rm lines} \langle N_{\rm b} \rangle$, with
$\langle N_{\rm b}\rangle$ being the number of sources in a telescope beam,
averaged over all directions.  The total fraction of sources contained
in the $\Nsurvey$ survey ``cells'' is given by 
$\epsilon_{\rm \survey} =
\epsilon_{\rm sky}\epsilon_{\rm ch}$.

The fraction $\epsilon_{\rm sky}$ can be related to the number density
of sources $n_{\rm D}$ defined in \S\ref{sec:demography} As we show in
that section,
even a small number of transmitting civilizations, $\approx 10^3$, can
result in more than one source per beam so that $\epsilon_{\rm sky}$
need not be small and may in fact approach unity.  However, it is
assumed that transmitting civilizations will use narrow bandwidths,
$\lesssim 1$ Hz, in a small number of individual channels.  Since the
center frequencies of such narrowband signals are not known {\em a
priori}, it is obviously necessary to cover as large a range in
frequency as possible in the survey, necessitating that the number of
channels, $N_{\rm ch}$, be quite large (e.g.,  in META, $N_{\rm ch}
\approx 10^7$).  Consequently, we expect 
$\epsilon_{\survey} \lesssim \epsilon_{\rm ch} \ll 1$.

Within the context of the three models presented in
\S\S\ref{sec:basic}--\ref{sec:obs_events}, the sampling pdf for the candidate
intensities is
\begin{eqnarray}
f_I^{({\rm \survey})}(I)
 & = & (1 - \epsilon_{\rm \survey} - \epsilon_{\rm RFI}) f_{I}(I, S=0) \nonumber \\
 & + & \epsilon_{\rm \survey} \int dS\,f_S(S) f_{I, \rm scint}(I, S) \nonumber \\
 & + & \epsilon_{\rm RFI} \int dS\,f_{\rm RFI}(S_{\rm RFI}) f_{I}(I, S_{\rm RFI}).
\label{eq:f_survey}
\end{eqnarray}
Here the first term accounts for observation cells 
devoid of sources, the second
term accounts for those cells with scintillating sources, and the
third term accounts for transient sources.  Analogous to
$\epsilon_{\rm \survey}$, we have introduced $\epsilon_{\rm RFI}$, the
fraction of observation cells in which there is a transient
signal.  Just as $\epsilon_{\rm\survey}$ can be related to the number
of transmitters in the Galaxy,
the rate at which transient signals occur and their duration can be
related to $\epsilon_{\rm RFI}$.

Following \S\ref{sec:obs_events} we use likelihood functions involving
$f_I^{({\rm \survey})}$ to test various models.  Our goal is to test
the hypothesis that a population of scintillating sources can account
for some of the events seen in SETI programs and to estimate
parameters of any such population.  Accordingly, we shall consider the same models as in
\S\ref{sec:obs_events}: a population of scintillating sources, Model~II, for which $\epsilon_{\rm RFI} = 0$, and an RFI-only explanation
for the events, Model~III, for which $\epsilon_{\rm \survey} = 0$.
However, should there be a method of distinguishing between
terrestrial RFI and narrowband celestial transients in the future, our
analysis can easily be extended to include such possibilities.

We shall also need the detection probability for the survey.
Integrating Eq.~\ref{eq:f_survey}, we have
\begin{equation}
P_{\rm d}^{(\survey)}
 = (1-\epsilon_{\survey}) P^{(\survey)}_{\rm fa}
 + \epsilon_{\survey} \int dS \, f_S(S) P_{\rm d,scint}\left(I^{(\survey)}_{\rm T}, S\right),
\label{eq:pd_survey}
\end{equation}
for Model~II; for Model~III, a similar expression holds with the
substitution of $\epsilon_{\rm RFI}$ for $\epsilon_{\rm \survey}$ and
$P_{\rm d}(I^{(\survey)}_{\rm T}, S)$ for $P_{\rm d,
scint}(I^{(\survey)}_{\rm T}, S)$.  Here, $P^{({\rm \survey})}_{\rm
fa}$ refers to the false-alarm probability for the survey as a whole,
hence, $\Nsurvey P^{(\survey)}_{\rm fa} \sim 1$.  We expect
\begin{equation}
\langle N_{\rm cand}^{(\survey)} \rangle 
 = N_{\rm su}  P_{\rm d}^{(\survey)}
\label{eq:nc_survey}
\end{equation}
to be the expected number of candidate signals found in the survey.
If there is noise only,  
amplitudes will follow the exponential pdf that results from
Eq.~\ref{eq:pdf} with $S=0$.


In practice, the noise level $\Nave$ is not constant over all sky
positions, due to ground spillover and Galactic latitude dependent
backgrounds.  A survey may be divided into regions where $\Nave$ is
constant and each region can be analyzed individually.
Here, however, since our goal is to exemplify a method of inference,
we assume $\Nave$ to be constant.

\subsection{Likelihood Function for the Entire Survey}\label{sec:like_survey}

The likelihood function for the survey is
\begin{equation}
{\cal L}^{({\rm \survey})}
 = {\cal L}_{\rm cand} {\cal L}_{\rm non},
\label{eq:like1}
\end{equation}
where the factor for survey candidates is
\begin{equation}
{\cal L}_{\rm cand} = \prod_{j=1}^{N_{\rm cand}} f_I^{({\rm
\survey})}(I_j)
\label{eq:like_c}
\end{equation}
and the non-detection factor is
\begin{eqnarray}
{\cal L}_{\rm non}
 & = & \prod_{j=1}^{\Nsurvey - N_{\rm cand}} P\left(I < I^{(\survey)}_{\rm T}\right)
 = \left [ 1 - P_{\rm d}^{(\survey)} \right ]^{\Nsurvey - N_{\rm cand}}.
\label{eq:like_non}
\end{eqnarray}
For our assumed form of $f_I^{({\rm \survey})}$, the parameters of
interest are the fraction of survey cells containing sources or RFI,
$\epsilon_{\rm \survey}$ and $\epsilon_{\rm RFI}$, and the parameters
of the source and RFI strength distributions, $f_S(S)$ and $f_{\rm
RFI}(S)$.  

As in our analysis of individual events in \S\ref{sec:like_events_u},
we again assume that reobservations occur at times such that the
scintillation gain is uncorrelated between all survey and
reobservation measurements\footnote{
{Assumption of uncorrelated vs. correlated reobservations is not
critical; reobservations of survey candidates are dominated by the
vastly greater number of nondetections in the survey itself.}
}.
With $K_{\rm r}(j)$ reobservations per candidate, 
the likelihood function for the reobservations is 
\begin{equation}
{\cal L}^{(\reobs)}
 = \prod_{j=1}^{N_{\rm cand}} \prod_{k=1}^{K_{\rm r}(j)} P\left(I< I^{(\reobs)}_{\rm T}\right)
 = \left [ 1 - P_{\rm d}^{(\reobs)} \right ]^{\sum_j^{N_{\rm cand}} K_{\rm r}(j)} .
\label{eq:like_ro}
\end{equation}
Equivalent to the survey detection probability in
Eq.~\ref{eq:pd_survey} is a reobservation detection probability
\begin{equation}
P_{\rm d}^{(\reobs)}
 = (1-\epsilon_{\reobs}) {P^{(\reobs)}_{\rm fa}}
 + \epsilon_{\reobs} \int dS \, f_S(S) P_{\rm d,scint}\left(I^{(\reobs)}_{\rm T}, S\right)
\label{eq:pd_re-obs}
\end{equation}
where we have distinguished, through the label ``{\reobs},'' the
false-alarm probability, the source-detection probability, the
threshold, and the fraction of cells containing sources in the
reobservations from those conducted in the survey observations.
Reobservations have a lower threshold intensity, $I^{(\reobs)}_{\rm T}$,  
and, accordingly, larger false-alarm and detection probabilities.
The beam fraction, $\epsilon^{(\reobs)}_{\rm sky}$, is also different from that
in the survey because directions are selected in the survey that have
shown strong signals.  If there is a population of sources we would
expect $\epsilon_{\rm sky}^{(\reobs)} \to 1$, but we still expect
$\epsilon_{\rm ch}^{(\reobs)} \ll 1$ and $\epsilon_{\reobs} \lesssim
\epsilon_{\rm ch}^{(\reobs)}$.  For instance, to cover a bandwidth
equivalent to the Doppler shift resulting from the Earth's motion,
$10^{-4}c$, at an observing frequency of 1~GHz requires $10^5$
channels of 1 Hz each.  In this example, therefore, $\epsilon_{\rm
ch}^{(\reobs)} \sim 10^{-5}$.

In our earlier analysis of the individual events,
\S\ref{sec:like_events_u}, we assumed that $S_{\rm RFI} = 0$ during
reobservations.  In contrast, since we are now analyzing the
{\it entire} survey,
we must allow for the possibility of RFI to be present during the
reobservations.  Consequently, in considering reobservations in
Model~III, we use Eq.~(\ref{eq:pd_re-obs}) with the substitution
$P_{\rm d}\left(I^{(\reobs)}_{\rm T}, S\right)$ in place of $P_{\rm
d,scint}\left(I^{(\reobs)}_{\rm T}, S\right)$.

The complete likelihood function is
\begin{equation}
\cl = \cl^{({\rm \survey})} \cl^{({\rm \reobs})}.
\label{eq:like_all}
\end{equation}
In practice, we analyze the log likelihood, 
\begin{equation}
\Lambda \equiv \log {\cl} = 
     \Lambda_{\rm cand} + \Lambda_{\rm non} +
\Lambda_{\rm \reobs}.
\label{eq:Lambda}
\end{equation}
Parameter values may be estimated through maximization of $\Lambda$.
Alternatively, Bayes' theorem may be applied by multiplying $\cl$ by a
{\em prior\/}  probability density for the parameters and normalizing to
obtain the {\em posterior\/}  pdf for the parameters.  In the following
we analyze only the likelihood function.  This choice is made because
(1) the pdfs for scintillation gain and source strength are already
built into the likelihood function; and (2) any {\em prior\/}  pdfs for
the relevant parameters, $\zeta_1$, $\zeta_0$, $\epsilon_{\survey}$,
and $\epsilon_{\reobs}$, would be so broad that they would not change the
net results.

\subsection{Fitting Model~II to META}\label{sec:meta3}

In Fig.~\ref{fig:meta_models}{\it a\/} and {\it b\/} we show the
likelihood functions for Model~II at 1420 and 2840~MHz, respectively,
using the META results (HS93).  
Table~3
reports the location and amplitude of the maximum in the likelihood
function for both the survey and the survey plus reobservations of
candidates.  In column~1 is the radio frequency, column~2 is the
maximum amplitude of the logarithm of the likelihood function, and
columns~3 and~4 are the maximum likelihood values for
$\zeta_0\epsilon_{\survey}$ and $\zeta_1$, respectively.  We do not
present the results for $\Lambda_{\rm cand}$ or $\Lambda_{\rm non}$
because there exists no single maximum for these functions, but rather
a region of constant and maximum likelihood.

There are several aspects of these results which deserve explanation.
Horowitz \& Sagan~(1993) report 14 candidates at 1420~MHz and 23
candidates at 2840~MHz, all with intensities greater than $28\Nave$.
However, they also acknowledge that they have included candidates just
below threshold so as not to exclude potential sources.  In producing
these likelihood functions, we have used only those 11 candidates
above the proper survey threshold, $\eta_{\rm T}^{(\survey)} = 31.7$.
We have also assumed that all reobservations were conducted with a
threshold of $\eta_{\rm T}^{(\reobs)} = 20$ ($P_{\rm fa} = 10^{-8.7}$)
and that each candidate was reobserved $K_j = 256$ times.  In order to
continue plotting the likelihood functions as functions of only two
parameters, we have set the ratio, $r_\epsilon \equiv
\epsilon_{\reobs}/\epsilon_{\survey}$, to be constant and equal to
unity.  Finally, the likelihood function for Model~II is plotted as a
function of $\zeta_0\epsilon_{\survey}$ and $\zeta_1$.  The degeneracy
between the parameters $\zeta_0$ and $\epsilon_{\survey}$ arises
because $f_S(S) \propto \zeta_0$, cf. Eqs.~\ref{eq:f_survey},
\ref{eq:pd_survey}, and \ref{eq:pd_re-obs}.  In the limit $\zeta_0 \ll
1$, the quantities $f_I^{(\survey)}(I)$, $P_{\rm d}^{(\survey)}$, and
$P_{\rm d}^{(\reobs)}$ become approximately linear in the quantity
$\zeta_0\epsilon_{\survey}$.

There is no difference in either the magnitude or location of the peak
likelihood when comparing $\Lambda_{\survey}$ and
$\Lambda_{\survey+\reobs}$.  The lack of the reobservations' influence
arises because of the vastly larger number of nondetections in the
original survey than reobservations.  For the non-detections,
\begin{equation}
\Lambda_{\rm non} 
 \approx \Nsurvey\log\left[1 - P_{\rm d}^{(\survey)}(\eta_{\rm T}^{(\survey)}, \epsilon_{\rm su}, \zeta_1)\right] 
 \approx -\Nsurvey P_{\rm d}^{(\survey)},
\end{equation}
where we have used the expansion $\log(1 - x) \approx -x$.  A similar
expression holds for $\Lambda_{\reobs}$, 
$\Lambda_{\reobs} \approx -N_{\rm cand}K_{\rm r}P_{\rm d}^{(\reobs)}$.
These two likelihood functions have a similar shape in the 
$(\zeta_0\epsilon_{\survey}, \zeta_1)$ plane.  However, because 
$\Nsurvey \approx 10^{13}$ while $N_{\rm cand}K_{\rm r} \approx 10^3$, 
the nondetection likelihood completely dominates the influence of
the reobservation likelihood.

Thus, $\Lambda_{\survey+\reobs}$ is dominated by $\Lambda_{\rm cand}$
and $\Lambda_{\rm non}$.  $\Lambda_{\rm cand}$ increases (and
$\Lambda_{\rm non}$ decreases) if either $\zeta_0\epsilon_{\survey}$
or $\zeta_1$ (or both) increases, as a result of the larger number of
sources that can be detected.  When these two are combined to produce
$\Lambda_{\survey+\reobs}$, the maximum of the likelihood function
occurs at values for $\zeta_0\epsilon_{\survey}$ and $\zeta_1$ that
are intermediate to those which respectively maximize $\Lambda_{\rm
cand}$ and $\Lambda_{\rm non}$.  As $S_1$ is the {\em upper cutoff\/}
to the source strength pdf, a simple constraint on $\zeta_1$ is that
it cannot be too much smaller than the intensity of the strongest
candidate observed\footnote{ { $\zeta_1$ can be smaller than the
intensity of an observed candidate by virtue of upwards noise and
scintillation fluctuations.}  }.  Similarly, given the small number of
candidates (i.e., a few tens out of $\sim 6 \times 10^{13}$ total
trials), we would expect that $\epsilon_{\survey}$ is small.  The
contours extending to very large $\zeta_1$ arise, in part, because of
the exponential pdf for the scintillation gain.  Since the most
probable gain is $g = 0$,
very large $\zeta_1$ can be tolerated and the absence of sources with
$\zeta \sim \zeta_1$ in the survey is explained by assuming that
scintillations modulated any such sources below threshold.

The locations of the peaks at the two frequencies compare favorably.
The location of the peak likelihood at 1420~MHz is within 10\% of the
peak likelihood at 2840~MHz and the 2840~MHz peak likelihood is within
50\% of the 1420~MHz peak likelihood.  The lack of better agreement
can be attributed to the fact that there were twice as many 2840~MHz
candidates as 1420~MHz candidates and the strongest 2840~MHz candidate
had $\eta = 746.6$ while the strongest 1420~MHz candidate had $\eta =
224$.

The difference in strongest candidate signals between the two
frequencies would seemingly explain the difference in the amplitude of
the likelihood functions as well.  It does not.  At 1420~MHz, there
are four candidates, three with $\eta \approx 34$ and one with $\eta =
224$.  At 2840~MHz, there are seven candidates, three with $\eta
\approx 32$, one at $\eta = 746.6$, and three with $40 < \eta < 80$.
It is these last three candidates which cause the marked difference in
the maximum value of the likelihood functions, simply because there
are so many of them.  An occasional large scintillation gain can
combine with a large noise fluctuation (cf. \S\ref{sec:meta1}) to
produce a signal well above threshold.  Hence, the likelihood function
is somewhat insensitive to one candidate with an observed intensity
well above threshold.  However, obtaining several, independent
combinations of such gains and noise fluctuations becomes
increasingly, and rapidly, less likely.

\subsection{Fitting Model~III to META}\label{sec:meta4}

The above discussion implicitly assumes that Model~II (a population of
scintillating sources) is the correct model.  To compare Model~II with
Model~III (the RFI model), we have also constructed the likelihood
functions for this model, Fig.~\ref{fig:meta_models}{\it c\/} and {\it
d}.  The maxima and locations of the likelihood functions are
tabulated in Table~3.

In estimating these likelihood functions, we have taken $f_{\rm RFI}$ 
to be a flat function, as in \S\ref{sec:bayes_events}, with an upper 
limit $S_{1,\rm RFI}$ and a lower limit $S_{0,\rm RFI} = 0$.  As for Model~II, we 
have assumed that each candidate position was reobserved 256 times and 
that $r_\epsilon $ is constant and equal to unity.

The likelihood functions for Model~III are very similar to those of
Model~II.  Again $\Lambda_{\survey+\reobs} \approx \Lambda_{\survey}$
because of the vastly larger number of survey observations;
$\Lambda_{\rm cand}$ increases with increasing $\epsilon_{\rm RFI}$
and/or $\zeta_{1,\rm RFI}$ while $\Lambda_{\rm non}$ decreases, with
the combination of these two producing a peak in $\Lambda_{\survey}$.

There are two notable differences between the likelihood functions for
the two models, both arising from the assumed $f_{\rm RFI}$.  Since
the pdf is flat, all intensities are equally likely, in contrast to
Model~II, where it is possible to ``hide'' very large intensities with
a very small scintillation gain.  Consequently, a smaller
$\epsilon_{\rm RFI}$ is required for Model~III than Model~II in order
that there not be too many candidates.  Similarly, the peak likelihood
tends to be more concentrated about the most likely $\zeta_{1,\rm
RFI}$, rather than extending to very large values as for Model~II.

The key issue between these two models, though, is the value of the
likelihood functions at the peak.  At both frequencies, we find that
the maximum likelihoods differ only slightly, with the RFI model
slightly preferred at 1420~MHz and the scintillating source model 
slightly preferred at 2840~MHz.  {\em We conclude that we are unable to
favor either Model~II or Model~III.}

\subsection{Interpretation of Model~II Fitting Results}\label{sec:mod2fit}

The fitting results yield estimates for $\zeta_0\epsilon_{\survey} =
\zeta_0\epsilon_{\rm sky}\epsilon_{\rm ch}$ and $\zeta_1$, the
normalized upper cutoff of the source flux pdf.  Using results from
\S\ref{sec:eti_transmit} and Eq.~(\ref{eq:esky})--(\ref{eq:ech}), we
may relate these estimated quantities to the population parameter
$N_{\rm D}$, the number of sources in the Galaxy.
Figure~\ref{fig:cls_epssky} shows $\epsilon_{\rm sky}$ as a function
of the mean free path (c.f. Eqs.~[\ref{eq:n_in_beam_mfp}] and
[\ref{eq:esky}]) for a simple disk model for the Galaxy.  We have
assumed a circular disk of radius $R_{\rm G} = 15$~kpc and thickness
$2H_{\rm G} = 0.2$~kpc with the Sun at the disk center
(heliocentricity for simplicity!).  From the fitting results we can
derive two estimates for $N_{\rm D}$, the number of sources in the
Galaxy, in terms of the survey parameter, $\epsilon_{\survey}$.  The
first relation results from inverting the relation $\epsilon_{\survey}
= \epsilon_{\rm sky} \epsilon_{\rm ch}$ where $\epsilon_{\rm sky}$ is
related to $D_{\rm mfp}$ (as in Fig.~\ref{fig:cls_epssky}) which,
in turn, is a function of $N_{\rm D}$, 
$D_{\rm mfp} \propto N_{\rm D}^{-1/3}$: 
\begin{equation}
N_{\rm D}^{(1)}
 = \epsilon_{\rm sky}^{-1} (\epsilon_{\survey} 
      N_{\rm ch} / P_{\nu}),
\label{eq:nd1}
\end{equation}
where
`$\epsilon_{\rm sky}^{-1}$' means the inverse of the function
$\epsilon_{\rm sky}(N_{\rm D})$ that relates 
the number of sources $N_{\rm D}$ to $\epsilon_{\rm sky}$;
 $P_{\nu}$ is the probability that a signal is in the band
observed by META.  Here we define this probability as
$P_{\nu} = N_{\rm signals/beam}\Delta\nu_{\rm spectrometer}/B_{\rm D}$. 

The second relation is based on the relationship between $N_{\rm D}$
and the flux ratio $S_1/S_0 \equiv \zeta_1/\zeta_0$.  We can determine
only $\zeta_1$ and the product $\zeta_0\epsilon_{\rm su}$ from the
likelihood analysis.  Let the ratio of these fitting parameters be
\begin{equation}
F \equiv \frac{\zeta_1}{\zeta_0\epsilon_{\rm su}}.
\label{eq:F}
\end{equation} 
Inverting the ratio $\zeta_1/\zeta_0$ (cf. \S\ref{sec:candles} yields
\begin{equation}
N_{\rm D}^{(2)} = 
\cases{
\frac 
     {\displaystyle F\epsilon_{\rm su}} 
     {\displaystyle 1 - F\epsilon_{\rm su}/2\Ntwid} & 
     $N_{\rm D}^{(2)}\le\Ntwid$\cr
\cr
(F\epsilon_{\rm su})^{3/2}  \Ntwid (H_{\rm G}/R_{\rm G})^3  & 
     $N_{\rm D}^{(2)} \ge\Ntwid$. \cr 
}
\label{eq:nd2}
\end{equation}

Figure~\ref{fig:cls_solutions} shows the two estimates for $N_{\rm D}$
plotted against $\epsilon_{\rm su}$ for $P_{\nu} = 0.01$.  Increasing
$P_{\nu}$ shifts $N_{\rm D}^{(1)}$ (solid curve) to the right.
Therefore, in order that there be a solution defined by the crossing
points of the two curves, $P_{\nu}$ must exceed a minimum of about
$0.003$.  This would suggest that the total radio domain of ETI
signals is about 300 times the META bandwidth if there is only one ETI
transmitter per beam.  But with more than one transmitter per beam,
the spectral domain can exceed the META bandwidth by much more than
this factor.

For values of $P_{\nu}$ that provide solutions, the number of civilizations
can be anywhere from a few to in excess of $10^{10}$.   The reason for
this indeterminacy is that, with larger $\epsilon_{\rm su}$, $\zeta_0$
decreases.  This implies, simply, that though there  are many more sources
for larger $\epsilon_{\rm su}$, the vast majority is
buried in the noise.  Unfortunately, the ambiguity in $N_{\rm D}$ means we 
cannot establish, through this analysis, 
 the distance scale for putative sources consistent with 
META candidates.  Therefore we cannot constrain the transmitter power
with the available information we have used. 

Horowitz \& Sagan point out that the strongest candidates in META
cluster about the Galactic plane.  We can use the
spatial distribution to estimate the distance scale and thus
provide another constraint on $N_{\rm D}$.   For the uniform disk
model of \S\ref{sec:eti_transmit}, and where sources can be
detected above threshold out to a distance $D_{\rm max}$ with
$H_{\rm G} \le D_{\rm max} \le R_{\rm G}$, 
we have
\begin{equation}
\langle \sin^2 b \rangle = 
   \frac{1}{3} \left ( \frac{H_{\rm G}}{D_{\rm max}} \right )^3
     \left ( 1 + 3 \ln \frac{D_{\rm max}}{H_{\rm G}} \right ).
\label{eq:sin2b}
\end{equation}
From the nine strongest META candidates, we
estimate $\langle \sin^2 b \rangle \approx 0.097$ compared to
a value $1/3$ for an isotropic population, 
implying $D_{\rm max} / H_{\rm G} \approx 2.3$.
The disk subvolume sampled by these nine events is 
$V_9 = \frac{2\pi}{3} H_{\rm G}^3 
  \left [ 3 (D_{\rm max}/H_{\rm G})^2 - 1 \right ]$
out of a total volume  $ V = 2\pi H_{\rm G} R_{\rm G}^2$.   For 
$R_{\rm G} / H_{\rm G} = 150$, as in \S\ref{sec:eti_transmit},
we find that $V/V_9 \approx 10^{3.7}$.  Thus the nine events found
in $V_9$ imply the presence of 
$N_{\rm D} \approx 10^{4.6}$ sources in the Galaxy. 
This number is comparable to the cross-over point of the two
lines in Figure \ref{fig:cls_solutions}.   

We conclude that
if the nine strongest META events are real celestial sources
(natural or artificial), there are a few tens of thousands of
such sources in the Galaxy.



\section{DUAL STATION \& DUAL BEAM OBSERVATIONS}\label{sec:bistatic}

Here we discuss the joint intensity statistics of SETI made 
simultaneously at two sites.  Most of our discussion also applies
to multiplying interferometry, such as proposed by 
Welch (1983).
Simultaneous SETI observations at two terrestrial locations are
analogous to pairs of single-site observations made at different times
when one considers the level of correlation for the interstellar
scintillations.  

In Appendix~\ref{app:stats} we discuss the spatial
correlation of ISS.  The correlation length is directly related to the
scintillation time because the underlying process is a diffraction
pattern swept across the line of sight at a transverse speed $\vperp$.
Referring to Fig.~\ref{fig:cls_iss}, scintillation times for nearby
sources are measured in hours while much faster scintillations, with
time scales of seconds, will be seen from sources across the Galaxy.
Transverse speeds $\vperp \sim 10$ km s$^{-1}$ then correspond to
length scales ranging from tens of km to $> 10^4$ km.  Dual-site SETI
observations will therefore encompass all possible degrees of
correlation of the scintillation gain.  These are (1) perfect
correlation for nearby sources; (2) partial correlation for
intermediate distances; and (3) no correlation for distant sources.
At 1.4~GHz, sources within 1~kpc will show highly correlated
scintillations for any pair of terrestrial sites and those at
distances greater than 5--8~kpc will display uncorrelated gains for
sites more than 1000 km apart.  We emphasize that the relevant
distances for demarcating these regimes are frequency dependent.  At
frequencies higher than 1.4~GHz, more sources in the Galaxy will show
scintillations that are spatially correlated between two sites.

Dual-site observations are therefore equivalent, statistically,  to 
single-site  reobservations 
made promptly after an initial detection (such that 
ISS is strongly correlated) and those made with substantial time delay,
in which case the ISS is uncorrelated.   As we saw in \S\ref{sec:pdfs},
the probability of detecting an ETI signal
in even a prompt reobservation need not be large and the 
probability of redetection
in a delayed, uncorrelated reobservation can be negligible. 


If $N_{\rm trials,ISS}$ is large (cf. Eq.~\ref{eq:trials_iss}), 
a survey is expected to encounter rare, high amplitude
scintillations, in which case weak sources may be detected at one site
 as a combination of noise and signal with low chances of redetection
(either later in time or at the other site).     
This conclusion restates our analysis of META events made in 
\S\ref{sec:like_events_u}.  Similarly, small $N_{\rm trials,ISS}$ means that
only smallish scintillation gains are to be expected in a survey.
In both cases, however, rare noise fluctuations play a key role
in detections of sources because of the large number of noise trials.

We refer to the analysis of \S\ref{sec:pdfs} on temporal sequences of
observations to analyze dual-site observations.   Figure~\ref{fig:cls_p2cond}
shows the detection probability of a second observation with
threshold $I_{\rm T}$ given that a first observation was made that yielded
an intensity $I_1$.  The detection probability is plotted against
the correlation coefficient $\rho$ (which may be either the temporal or
the spatial correlation).  The results for the several values of
$I_{\rm T}$ and $0 \le \rho \le 1$ indicate that the probability of 
detection at a second site is small, even when $\rho$ is large.     

\subsection{META \& META~II Two-station Observations}\label{sec:metaii2}

META and META~II devote some observing time to simultaneous
observations at 1.42~GHz and with a survey threshold $\eta_{\rm T} =
16$ at each site.  For detections at threshold at one site,
i.e., $\eta_1 = 16$, Fig.~\ref{fig:cls_p2cond} provides the detection
probability for the other site.  Even for large intrinsic signal
strengths, e.g., $\zeta = 32$, the chance of {\em not\/}  detecting the
source at the second site is non-negligible, $\approx 35$\%.  If the
initial detection was of a weak source, e.g., $\zeta \le 4$, modulated
above threshold by a large scintillation gain or noise fluctuation or
both, the probability of detecting the source at the second site can
easily be less than 10\%.  In accord with our previous discussion, a
significantly lower threshold (e.g., $10\Nave$,
c.f. Fig.~\ref{fig:cls_p2cond} and \S\ref{sec:pdfclean}) would make
detections at both sites much more probable, albeit with a larger
background false-alarm rate.  Part of the task of subsequent
processing would be an assessment of whether excess hits above the
background rate were statistically significant (see \S\ref{sec:pdfclean}).

\subsection{Future Two-station or Dual-Beam Observations}
\label{sec:future_bistatic}

The recommended procedure for dual-site observations is to report
observations at low thresholds.  This necessitates a large amount of
storage for results in a dual-site search but raises the odds of
a dual detection.  As discussed in \S\ref{sec:pdfclean}, 
there are additional reasons for reporting more measurements than are
typically done in SETI.

In the proposed BETA survey, a billion channel spectrometer will be
used in a two-feed antenna system so that a given sky position is
viewed successively in the two beams (HS93).  The time delay is
of order minutes, so scintillating sources may or may not be
correlated in the two beam measurements, depending on observation
frequency and source distance.  The odds for dual detections depend on
the observing threshold according to our previous considerations, with
lower thresholds preferred.

\subsection{SETI Interferometry}\label{sec:seti-int}

Welch (1983) has emphasized the inferential power of multiplying
interferometry in SETI with respect to localization on the sky
and for establishing the celestial nature of any real sources
by seeing the influence of the expected Doppler shifts.  ISS
influences interferometric observations by attenuating the visibility
function by a factor $\rho_s(b)$, where
$b$ is the interferometer baseline.

\section{ALTERNATIVE DETECTION METHODS}\label{sec:pdfclean}

The conventional detection method in SETI and other astronomical surveys
uses a threshold in signal-to-noise ratio (SNR) to define candidate
signals.  The false-alarm probability is often used to define the threshold 
in terms of the number of statistical trials made in the survey. 

Here we advocate an alternative method for SETI that uses more
information in the data and {\em a priori\/}  knowledge about the noise
statistics.  This method allows lower signal levels to be probed than
in the SNR method.  The scheme, first discussed by Zepka, Cordes \&
Wasserman~(1994) for detecting X-ray sources in Poisson backgrounds,
looks for departures from the expected shape of the intensity (or
count-rate) histogram.  The stopping criterion is that sources are
identified iteratively until the histogram of residual 
intensities is consistent
with a noise-only histogram.  For this reason, we refer to the method
as ``pdfCLEAN.''

For interference-free and source-free observations, the {\em a
priori\/} intensity pdf is given in Eq.~\ref{eq:pdf} with $S=0$,
and is a one-sided exponential, $\exp(-\eta)$.
The histogram of intensities, $h_k$, is the number of counts in the
$k^{\rm th}$ intensity bin with centroid value $\eta_k$.  
For $N_{\rm obs}$ 
total observations, the mean value of the histogram is
\begin{equation}
\langle h_k \rangle = N_{\rm obs} \exp(-\eta_k).
\label{eq:pdfclean1}
\end{equation}
The {\it actual} number of counts in $h_k$ is 
a Poisson random variable such
that the probability of obtaining $h_k$ counts is given by
\begin{equation}
P(h_k)
 = \frac{{\rm e}^{-\langle h_k \rangle} \langle h_k \rangle^{h_k} }{h_k!}.
\label{eq:pdfclean2}
\end{equation}
Departures from the expected exponential 
shape are identified by
seeking histogram bins where $P(h_k)$ is small, 
indicating an excess
or deficit of counts.  The pdfCLEAN method 
therefore makes use of a
{\it probability threshold} in place of an 
SNR threshold in the SNR method.
Once histogram bins are found in this way, they may be
``deconstructed'' to find the original sky positions and frequency
channels of the contributing intensities.  These may then be subjected
to clustering tests to see if real celestial signals or terrestrial
interference is the cause for departure.

Application of pdfCLEAN requires substantial storage for raw data
results because it can be used to probe intensities at lower SNR than
can the SNR method itself.  A discussion of actual storage
requirements is given below.  The payoff, however, is that real
sources can be identified at low levels and the interference
environment can be better understood as well.  Zepka et al.~(1994)
show that the number of spurious detections (e.g., due to noise
fluctuations) is small using pdfCLEAN.

\section{RECOMMENDATIONS FOR FUTURE SETI PROGRAMS}\label{sec:future}

Several lines of argument that we have presented lead to a primary
recommendation for future SETI: {\em Signal detections should be
reported at much lower signal levels than have been used}.  In
programs like META, where a large-scale survey is performed at one
threshold with subsequent reobservations at a lower threhold, we argue
that the reobservation threshold should be low enough to rule out our
Model~II where survey candidates are hypothesized to be scintillating
but otherwise constant sources.


The cost of reporting lower signal levels is a larger recording rate
and, possibly, a problematic level of false alarms.  The latter is
particularly the case if the candidate frequency is expected to vary
between survey detection and reobservation, necessitating a large
number of frequency channels to be searched during the reobservations.
However, in \S\ref{sec:pdfclean} we argue that false-alarms need not
deter a survey from detecting weak sources.  Use of the Zepka et
al.~(1994) pdfCLEAN method allows signals to be extracted from
histogram counts that contain many noise-only measurements (i.e., false
alarms).

To estimate the threshold to be used for recording spectral amplitudes,
we consider the technological limits
on practical data recording.  At present, it is feasible to record approximately
one high-density magnetic tape per day and expect to analyze it at approximately
the real-time rate on a network of workstations.  
This is approximately 10 Gbytes/day.   
The data rate is
\begin{equation}
\dot\beta_{\rm max}
 = \frac{{\rm e}^{-\eta_{\rm T}} N_{\rm ch}\beta}{T_{\rm s}},
\label{eq:hitrate}
\end{equation}
and the corresponding threshold is
\begin{equation}
\eta_{\rm T} = \ln \left [ \frac{N_{\rm ch}\beta}{\dot\beta_{\rm max} T_{\rm s}} \right ]
  = 9.5 + \ln \left [
                 \left (\frac{N_{\rm ch}}{10^9}\right ) 
                 \left (\frac{\beta}{16\,{\rm B}} \right )
                 \left (\frac{10 \, {\rm GB/day}} {\dot\beta_{\rm max}} \right )
                 \left (\frac{10\,{\rm s}} {T_{\rm s}} \right ) 
                 \right ],
\label{eq:hitthresh}
\end{equation}
where $T_{\rm s}$ is the integration time for one spectrum, $\beta$ is
the number of bytes recorded per threshold crossing or ``hit,'' and
$\dot\beta_{\rm max}$ is the maximum data recording rate.
In Eq.~\ref{eq:hitthresh} we have adopted parameters that typify
near-future SETI spectrometers having $\sim 10^9$ channels, such as
BETA (HS93) and SERENDIP~IV (\cite{bow94}).  The $\sim 16$
bytes recorded per hit would give the sky position, time, frequency
channel, signal amplitude, and some modest line-shape parameterization
(e.g., whether it was unresolved, its width, etc.)

A semi-quantitative measure of the improvement offered by a lower
threshold can be seen in Fig.~\ref{fig:cls_p2cond}.  With a threshold
of $\eta_{\rm T} = 10$, the detection probability $P_{\rm 2d}$ for an
event at the survey threshold of META, $\eta_1 = 32$, is nearly a
factor of $10^2$ greater than with the threshold used.  Similarly,
dual-site observations employing thresholds of 10 rather than 16,
e.g., META and META~II, would have larger detection probabilities by a
factor of $\sim 10$.

Whether a threshold as small as estimated is practical depends also on
the extent to which spectral intensities depart from exponential
statistics due to RFI, soft RAM errors, or celestial signals.  From
our considerations in previous sections, however, the hit threshold
presented here would be more than adequate for a full statistical
study of event amplitudes, especially if followed up with
reobservations having a smaller number of channels (by a factor of
$10^3$, say) and with a smaller threshold, $\eta_{\rm T}^{(\reobs)}
\sim 5$.  The data rate for reobservations would be substantially
smaller than in the survey and would therefore not have an impact on
the overall data rate.  Follow-up analysis would include rejection of
signals that appear in multiple sky positions and a search for signals
that repeat in single sky positions (consistent with the telescope
beam).  Application of the Zepka et al.~(1994) pdfCLEAN method would
search for intensity histogram bins which have an excess of counts.  In
this regard, pdfCLEAN can investigate signals at levels where noise
produces a nonnegligible number of false-alarm threshold crossings.

The threshold in Eq.~\ref{eq:hitthresh} is also adequate for two-station SETI, 
which can tolerate half the threshold of single-station SETI for roughly the 
same false-alarm rate.

For
intensities below the hit threshold, we advocate computation of a histogram of 
S/N  for a generous number of subbands of frequency channels and for a set of 
individual sky positions.   This low-intensity histogram provides the means for
identifying low-level RFI that persists in specific frequency bands and allows
redundancy testing for signals that repeat in more than one sky position. 

In summary, for future SETI  we advocate a survey threshold given by
$\eta_{\rm T}$ in Eq.~\ref{eq:hitthresh} and 
where a modest amount of information
is stored per hit (e.g., 16 bytes).  
Signals below the survey threshold would be recorded only
as a count in a histogram calculated for a given subband of frequency channels
and a coarse sky position.

\section{SUMMARY AND CONCLUSIONS}\label{sec:summary}

In this paper we have discussed how interstellar  scintillations cause
intermittency in SETI and we have developed a formalism based on
both the likelihood function and a Bayesian analysis for
analyzing existing and future surveys.  We emphasize that our results
are based on the case where interstellar scintillations are saturated,
producing modulations with a one-sided exponential distribution.
This situation applies for distances beyond a frequency-dependent
minimum distance, approximately a few hundred parsecs at 1 GHz.
Our conclusions do not apply, quantitatively, for SETI that targets
nearby stars using centimetric wavelengths, as in Project
Phoenix (\cite{t94}).  However, scintillation effects from the
solar wind and from the stellar wind of the host star
of transmitting sites will influence search sensitivities.  These
situations can also be analyzed using the methods of this paper. 

In applying our methods to existing surveys we have found that:

\begin{itemize}

\item In META (\cite{hs93}), considering each candidate
separately, it is extremely unlikely that any of the 9 candidates
with amplitude $\ge 33\sigma$ was a mere noise
fluctuation.  A real signal, terrestrial or extraterrestrial,
must underly these events.
  However, we are unable to distinguish between a steady signal
modulated above threshold by scintillations and a transient signal,
such as RFI,  soft RAM errors, or transient ETI signals themselves.

\item If any of the events in META is a due to a 
scintillating celestial source, 
it likely results from  a modest to large scintillation
gain {\em combined\/}  with a favorable noise fluctuation.

\item We also show that existing reobservations of META candidate signals 
(i.e., those performed to date) 
are incapable of ruling out the case where a real ETI source with
constant, intrinsic signal strength underlies the measured candidate
signal.  This conclusion holds even for the case where the scintillations
remain correlated between the time of an initial detection and prompt
reobservations.  Future reobservations are certainly capable of ruling
out a constant-source model for META detections.    

\item A stronger test of our signal models and of the celestial nature of
candidate signals requires much lower thresholds and a larger number of
reobservations than have been performed to date.  For META,
a reobservation threshold of 5--10 (in signal-to-noise ratio) rather
than the actual 20 is needed.  The number of reobservations needed is
a function of threshold but is many thousands for the threshold used.
In future surveys, correlated reobservations made promptly after the
initial survey detection can drastically lower the total number of
reobservations needed to rule out constant-strength ETI signal
hypotheses.

\item Dual-site observations made simultaneously are unlikely to 
yield dual detections of a source unless ``low'' thresholds are
used for recording signal levels.   

\end{itemize}

Having developed these formalisms and applied them to existing
surveys, we also present recommendations for planning future surveys.
An important recommendation concerns the philosophy regarding events
below threshold.  In Horowitz \& Sagan~(1993), nearly all events below
threshold were discarded.  As a result, in this paper, we have treated
all such events equally, since the only information reported was that
the intensity was below threshold.  In doing so, we have nevertheless
illustrated that, with the appropriate combination of scintillations
and noise, observed intensities of, say, $32\langle N\rangle$ can be
explained by scintillating sources.  If such scintillating sources do
exist, then there must be many occasions in which the intensity would
have been below threshold.  We recommend the following strategies,
based upon the methods described in Zepka, Cordes, \&
Wasserman~(1994), for establishing the existence of and
recovering these signals:

\begin{itemize}

\item{} A minimalist strategy would retain only
those events exceeding a threshold (defined by the usual false-alarm
probability), but would form a histogram of those events 
below threshold.
Since the expected distribution of the noise is known, viz. a
one-sided exponential, then {\em deviations\/} from this histogram
indicate the presence of non-noise signals.  Of course, 
separating RFI from actual ETI signals still remains a problem, 
requiring new observations.

\item{} The optimal strategy would 
retain individual intensity values  even for those
that are far below the false-alarm threshold.  
Clustering on the sky of results from multiple scans of the sky
would suggest the locations of extraterrestrial sources that could
be subjected to intense scrutiny by the flotilla of existing astronomical
instrumentation.

\end{itemize}

Application of our low-threshold strategy requires archiving of spectral
amplitudes at large data rates.   Our suggestion is that, given the 
large expense for conducting SETI, the data recorded and archived should
be of commensurate cost and at a volume dictated by hardware
available in the current market.

\acknowledgments 
We thank P. Horowitz, T. Loredo, J. Tarter, and D. Werthimer for
helpful discussions.  We also thank J. Tarter for organizing a
workshop on intermittency in SETI, co-chaired by Carl Sagan, and
co-sponsored by the Planetary Society, that was held at the SETI
Institute in 1994 January.  We thank the referee, B. Rickett,
for comments that helped clarify the paper.
Finally, JMC and TJWL thank Carl Sagan for decades of inspiration
in general and for his enthusiasm and clarity on the subject matter
of this article in particular.   
His last comment to us about the paper was that
``...it may very well lead to the ultimate success of SETI.''
  This work was supported by the National
Astronomy and Ionosphere Center, which operates the Arecibo
Observatory under a cooperative agreement with the NSF, and by the
Planetary Society,
\newpage


\appendix

\section*{APPENDICES}
\section{Symbols Used}\label{app:syms}
\begin{center}
\small
\begin{tabular}{ll}
\hline
\hline
%
Symbol & Definition \\
\hline

I,II,III &  Denotes Model~I, II, or III. \\ 

$\alpha$ &  The exponent in the $\log N$-$\log S$ distribution of 
		transmitters. \\ 

$B_{\rm D}$ &  Spectral domain in which ETI signals are
		transmitted. \\ 

$\beta$ & The number of bytes recorded per threshold crossing, or
	``hit.'' \\ 

$\dot\beta_{\rm max}$ & The maximum data recording rate in 
	bytes s$^{-1}$. \\ 

${\cal C}$ &  The set of candidate signals found in a survey. \\ 

\creobs  & Denotes reobservations with correlated scintillations.  \\ 

$D_{\rm mfp}$ & Mean free path for the line of sight intersecting a source.
		\\ 

$\epsilon_{\rm ch}$ & Fraction of spectrometer channels containing 
                ETI signals. \\ 

$\epsilon_{\rm RFI}$ &  Fraction of all observations that include 
   terrestrial RFI signals. \\ 

$\epsilon_{\rm sky}$ &  Fraction of telescope beams 
                  containing ETI sources. \\ 

$\eta \equiv I/\Nave$ &  The intensity measured in units 
                         of the mean background noise. \\ 

$\eta_{\rm c} \equiv I/\Nave$ &  The normalized intensity of a 
				candidate. \\ 

$\eta_{\rm T} \equiv I_{\rm T}/\Nave$ & The intensity threshold for 
				detection. \\ 
 
$f(\eta\vert\cal I)$ & The {\em posterior}\ pdf for the measured,
        normalized intensity \\
        & given the background information $\cal I$. \\ 
 
$f(\mbox{\boldmath $\theta$}\vert\eta{\cal I}) $ & {\em Posterior}\ pdf for
a set of parameters  \mbox{\boldmath$\theta$} given measurements \\
   & of the normalized  intensity $\eta$ and
   background information $\cal I$. \\ 

$f_{\rm I,II,III}$ & The {\em posterior}\ pdf for Model~I, II, or III. \\ 
ded
 
$f_g(g)$ &  The pdf of the scintillation modulation (or gain),
                $g$. \\ 
 
$f_{2g}(g_1, g_2; \rho)$ & The bivariate pdf for two values of
                          the scintillation gain $g_{1,2}$ \\
           & as a function of the correlation coefficent $\rho$. \\ 
 
$f_I(I,S)$ & Intensity pdf for a constant source strength. \\ 

\hline 
\end{tabular}

\small
\begin{tabular}{ll}
\\
\hline
\hline
%
Symbol & Definition \\
\hline

$f_{I,\rm scint}(I,S)$ & Intensity pdf for a source that varies due to 
                     scintillations. \\ 

$f_{2I}(I_1, I_2; S, \rho)$ &  The bivariate intensity pdf. \\ 

$f_{\rm RFI}(\zeta_{\rm RFI})$ & The pdf of normalized RFI source
		strengths. \\ 

$f_S(S)$ & The pdf for the intrinsic source strength of ETI
		transmitters. \\ 

$f_{\mbox{\boldmath{$\theta$}}}(\mbox{\boldmath{$\theta$}})$ & 
	The {\em a priori}\ pdf for a
	set of parameters \mbox{\boldmath{$\theta$}}. \\ 

$f_{\zeta} (\zeta)$ & Pdf of normalized source strengths based 
                     on the spatial \\
                    & distribution and intrinsic radiated 
                     powers of ETI sources. \\ 

$g$ &  Scintillation ``gain'' for a compact source's radio flux. \\ 

$g_{\rm c}$ &  Scintillation gain applicable to a 
               particular candidate event. \\ 

$H_{\rm G}$ &  Half-thickness of the galactic disk. \\ 

${\cal I}$ &  Background information in Bayesian analysis. \\ 

$I$ & Intensity. \\ 

ISS  & Interstellar scintillation or scattering. \\ 

$\delta I$ &   The rms variation in intensity. \\ 

$\Delta I$  & The uncertainty in establishing intensities during a
		survey. \\ 

$I_0(x)$ & The modified Bessel function of order $0$. \\ 

$I_{\rm cand}$ & Intensity of a candidate from a search program. \\ 

$I_{\rm T}$ &  Intensity threshold in a search program. \\ 

$K_{\rm r}$ & Number of reobservations of a candidate signal. 
             \\ 

$\krc$ & Number of reobservations where the 
         ISS gain is {\em correlated} \\ 
         &with the gain for the original candidate signal. \\ 

$\kru$ & Number of reobservations where the ISS  
          gain is {\it uncorrelated} \\
         & with that for the original candidate detection. \\ 
\hline 
\end{tabular}


\small
\begin{tabular}{ll}
\\
\hline
\hline
%
Symbol & Definition \\
\hline

$\krccrit$ &  Critical number of correlated reobservations needed to exclude \\
              & the reality of survey candidate \\ 

$\krucrit$ &  Critical number of uncorrelated reobservations needed to 
              exclude \\ 
              & the reality of a survey candidate. \\ 

$K_{\rm s}$ &  Number of separate observations of each 
               sky position in a survey. \\ 

${\cal L}$ & Likelihood function for a candidate signal. \\ 

${\cal L}_{\rm cand}$ & Likelihood function for the candidates
		from a survey. \\ 

${\cal L}_{\rm non}$ & Likelihood function for the non-detections
		in a survey. \\ 

$\lambda \equiv \log \cal L$ & Log likelihood function for an individual 
                              candidate signal. \\ 

$\Lambda$ &  Total log likelihood function for an entire
		survey. \\ 


$\ell_{\rm d}$ & Characteristic length scale of the diffraction
	pattern. \\ 

$\ell_{\rm D}$ & Typical distance between transmitters. \\ 

$\ell_{\rm min,max}$ &  Maximum and minimum distances 
               of Galactic transmitters. \\ 

${\cal N}$ & The set of nondetections in a survey. \\ 

$N$ & Noise intensity. \\ 

$\Nave$ & Ensemble average noise intensity. \\ 

$N_{\rm b}$ &  Number of Galactic transmitters in a 
		 telescope beam. \\ 

$N_{\rm cand}$ & Number of candidate signals found in a survey. \\ 

$N_{\rm ch}$  & Number of frequency channels in a SETI
		spectrometer. \\ 

$n_{\rm D}$ &  Number {\em density}\ of transmitters. \\ 

$N_{\rm D}$ &  Number of ETI  transmitters in the Galaxy. \\ 

$\Ntwid$ & Number of sources such that 
		$\ell_{\rm D} = H_{\rm G}$. \\ 

$\Delta N_{\rm fa}$ & Number of false alarms expected  
		 per frequency channel. \\ 

$N_{\rm frames}$ & Number of reference frames assumed for Doppler shifts. \\ 

\hline 
\end{tabular}

\small
\begin{tabular}{ll}
\\
\hline
\hline
%
Symbol & Definition \\
\hline

$N_{\rm lines}$ & Number of distinct spectral lines transmitted 
             by an ETI source. \\ 

$N_{\nu}$ &  Number of separate center frequencies searched. \\ 

$N_{\rm pol}$  & Number of polarizations searched. \\ 

$N_{\rm signals/beam}$ & Number of spectral lines from
	all ETI  sources in a telescope beam. \\ 

$N_{\rm sky}$ & Number of sky positions (telescope beam areas) searched. \\ 

$\Nsurvey = N_{\rm sky}N_{\rm ch}K_{\rm s}$ & Total number of statistical
		trials in a survey. \\  

$N_{\rm trials}$ & Total number of statistical trials in a 
	survey. \\ 

$N_{\rm trials,noise}$ & Number of trials of the noise in a survey. \\ 

$N_{\rm trials,ISS}$ &  Number of trials of independent scintillations  
                    in a survey. \\ 

$\Delta \nu_{\rm d}$ &  Characteristic ``diffraction'' 
                  bandwidth for scintillations. \\ 


$\Delta\nu_{\rm sb}$ & Characteristic broadening width for a
			spectral line. \\ 

$\Delta\nu_{\rm spectrometer}$ & Total bandwidth of a SETI
		spectrometer. \\ 

$\Omega_{\rm b}$ & Solid angle of a radio telescope's primary
		beam. \\ 

${\cal P}$ & Power radiated by a transmitter, assumed to be  
          radiated isotropically \\
           & [effective  isotropic radiated power (EIRP)]. \\ 

$P(I)$   &   Probability that the intensity exceeds a value 
		$I$. \\ 

$P_g(g)$ &   Probability that the scintillation gain exceeds a value
              $g$. \\ 

$P_{\rm d}(I_{\rm T},S)$ & Detection probability that a source with 
		strength $S$ \\
             & produces an intensity that exceeds a 
		threshold $I_{\rm T}$. \\ 

$P_{\rm d,scint}(I_{\rm T}, S)$ & Detection probability when a source
				of strength $S$ scintillates. \\ 

$P_{\rm 2d}(I_{\rm T} \vert I_1; S, \rho)$ 
       &  Detection probability in a second observation 
          given an initial detection \\
       &  with amplitude $I_1$, a source strength $S$, and 
            scintillation correlation $\rho$. \\ 

$P_{\rm fa}$ & Probability of a false-alarm detection. \\ 

$P_{\nu}$ &  Probability that a band of 
             frequencies contains ETI signals. \\ 


pdf      & Probability density function  \\ 
\hline 
\end{tabular}

\small
\begin{tabular}{ll}
\\
\hline
\hline
%
Symbol & Definition \\
\hline

${\cal R}$ & Set of intensity upper limits in reobservations 
         made on survey candidates.  \\ 

$r \equiv \cl_{\rm II}/\cl_{\rm III}$ & Ratio of likelihood
		functions for 
		Model~II to Model~III. \\ 

$r_\epsilon \equiv \epsilon_\reobs/\epsilon_\survey$ & 
		Ratio of the fraction of pointings containing 
		ETI transmitters for  \\
            & reobservations and survey. \\ 

$R_{\rm G}$ & Radius of the galactic disk. \\ 

$\rho_g(\tau)$ & Normalized, temporal  correlation function 
                 for the scintillation gain $g$  \\ 
               & as a function of time lag $\tau$. \\ 

$\rho_I(\tau)$ & Normalized correlation function for the
		intensity. \\ 

$\rho_N$ &  Normalized correlation function for radiometer noise. \\

$\rho_{\rm s}(\tau)$ &  Normalized spatial correlation function 
                for the scintillation gain $g$. \\ 

\reobs  & Subscript or superscript to denote 
		 reobservations of candidates. \\ 

$S$ &  Intensity of ETI signal without any modification by
	scintillations.\\ 

$S_{1,2}$ &  Minimum \& maximum fluxes of a galactic 
             population of standard candles.  \\ 

$S_{\rm RFI}$ & Intensity of a terrestrial RFI signal. \\ 

$S_{1,\rm RFI}$ & Maximum intensity of a terrestrial RFI signal. \\ 

$S_{0,\rm RFI}$ & Minimum Intensity of a terrestrial RFI signal. \\ 

\survey  & Subscript or superscript for a quantity relevant to a
		survey. \\ 

$\Delta t_{\rm d}$  & Characteristic diffraction time scale 
                for scintillations (strong scattering regime). \\ 


$T_{\rm s}$ & Integration time for one spectrum. \\ 


$\theta_{\rm b}$ &  One dimensional beam width (FWHM) of a 
           radio telescope's primary beam. \\ 

\ureobs  & Subscript or superscript to denote  
		 {\em scintillation-uncorrelated}\ reobservations \\ 

$V_{\rm b}$ &  Volume in the Galactic disk within a
		telescope beam. \\ 

$V_{\perp}$ &  Perpendicular speed by 
            which the observing geometry to a source changes. \\ 
\hline 

\end{tabular}

\small
\begin{tabular}{ll}
\\
\hline
\hline
%
Symbol & Definition \\
\hline

$\zeta \equiv S/\Nave$ & Source signal strength in units 
                of the mean background noise. \\ 

$\zeta_{1,2}$ &  Minimum and maximum normalized fluxes of a galactic \\ 
          &  population of standard candle transmitters. \\ 

$\zeta_{(0,1),\rm RFI}$ &  Minimum and maximum normalized RFI flux. \\ 

$\zeta_{\rm RFI}$ & RFI signal strength in units  
                        of the mean background noise. \\ 
\hline 

\end{tabular}

\end{center}

\newpage

\setcounter{equation}{0}
\section{Intensity Statistics}\label{app:stats}

Let the measured intensity (in a single channel of a spectrometer,
say) be
\begin{equation} 
I = \vert s + n \vert^2
\label{eq:I}
\end{equation}
where $n$ is complex, gaussian noise and $s$ is a signal phasor of {\em fixed\/}
amplitude.  The mean intensity is 
\begin{equation}
\langle I \rangle = S + \langle N \rangle
\label{eq:mean}
\end{equation}
where $S\equiv \vert s \vert^2$,  $N\equiv \vert n \vert^2$, and the variance
is
\begin{equation}
\sigma_I^2 = \sigma_N^2 + 2S\Nave.
\label{eq:variance}
\end{equation}
For unsmoothed noise from a Fourier transform spectrometer,
which we assume for the entirety of this paper,
$\sigma_N = \Nave$. 

The probability density function (pdf) of $I$ for an intrinsic
signal intensity $S$  is
\begin{equation}
f_{I}(I; S) 
  = \langle N \rangle^{-1} I_0\left (\frac{2\sqrt{IS}}{\langle N \rangle} \right )
       \exp\left[-\frac{(I+S)}{\langle N \rangle} \right ] U(I),
\label{eq:pdf}
\end{equation}
where $I_0$ is the modified Bessel function and $U(I)$ is the unit
step function (\cite{goo84}).  With no signal ($S=0$), the pdf becomes
a one-sided exponential function.  The pdf of $\sqrt I$ is the Rice
distribution (e.g., \cite{pap91}).

The realistic case, where the signal strength varies due to scintillations,
requires that $S\to gS$ in Eq.~\ref{eq:pdf} where $g$ is the appropriate
scintillation ``gain.''  Over short time spans (much less than a characteristic
scintillation time), $g = $ constant and the pdf of the intensity is of the 
form of Eq.~\ref{eq:pdf}.  However, over long time spans, $g$ varies according 
to its own pdf, $f_g(g)$, and the resultant intensity pdf is obtained by
integrating over the pdf of $g$:
\begin{equation}
f_{I,\rm scint}(I;S) = \int dg\, f_g(g)\, f_{I}(I; gS).
\label{eq:pdf_s}
\end{equation}
For saturated scintillations,
\begin{equation}
f_g(g) = \exp(-g)U(g),
\label{eq:pdf_g}
\end{equation}
yielding
\begin{equation}
f_{I, \rm scint}(I; S) 
 = \langle I \rangle^{-1} \exp\left(-\frac{I}{\langle I \rangle} \right)U(I), 
\label{eq:pdf_sat}
\end{equation}
where $\langle I \rangle = S + \Nave$, as before.

The utility of the two pdf's in Eqs.~\ref{eq:pdf} and \ref{eq:pdf_sat}
is as follows.  In surveys where the dwell time per sky position is
much less than the scintillation time scale, the apparent source
strength, $gS$, is fixed, and Eq.~\ref{eq:pdf} is the appropriate form
of the pdf to use in calculating detection probabilities, etc.  {\em
of that source at that time}.  In large scale surveys of many
independent sky positions, candidate detections of real signals will
preferentially select large values of the product $gS$ and, most
likely, will be predisposed toward selecting abnormally large values
of the scintillation gain $g$.  However, in reobservations of
candidate signals selected from an initial survey, it is likely that
the reobservations will span, {\em in toto}, many characteristic
scintillation times.  The scintillation gain for any real source will
vary over its allowed domain according to the exponential pdf (for
strong scattering).  In assessing the detectability of the
source in these multiple reobservations, one must use the second form
of the intensity pdf, i.e.,  Eq.~\ref{eq:pdf_sat}.

Throughout this paper, we assume that the strong scattering regime
obtains.  For completeness, we note that there are two other
scattering regimes (LC98):  The weak scintillation
regime is marked by a nearly symmetric scintillation gain pdf with a
considerably smaller variance and a much longer scintillation time
scale ($\sim 1$ day) than the strong scattering case.  In the
transition regime the gain pdf has no simple analytical form but, when
compared to an exponential distribution, has a much longer tail to
large $g$ while having a much longer scintillation time (of order the
time scale in weak scintillation).  At 1 GHz, the weak scintillation
regime extends to approximately 100 pc and the transition regime
extends to approximately 500 pc.  In order that there be sources
within these two regimes, the number of civilizations in the Galaxy
must exceed $10^5$ and $10^3$ for the weak and transition regimes,
respectively.  These numbers follow by simply calculating the 
$N_{\rm D}$ needed to make the mean distance between civilizations
equal to 100 or 500 pc.

\subsection{Signal Detection and False-alarm Probabilities}\label{sec:pd}

To calculate the probability that the intensity exceeds a 
specified threshold, $I_{\rm T}$, i.e., a ``detection,''
we integrate the appropriate pdf,
\begin{equation}
P_{\rm d}(I_{\rm T}; S) = P(I>I_{\rm T}) 
 =  \int_{I_{\rm T}}^{\infty} dI\, f_{I}(I;S),
\label{eq:pd_nonscint}
\end{equation}
for the case of a non-scintillating signal of strength $S$ or a single 
observation of a scintillating signal of observed strength $S \to gS$, 
and 
\begin{equation}
P_{\rm d,scint}(I_{\rm T}; S) = P(I>I_{\rm T}) 
 = \int_{I_{\rm T}}^{\infty} dI\, f_{I, \rm scint}(I; S),
\label{eq:pd_scint}
\end{equation}
for the case of many observations of a scintillating source.

For a scintillating source, the detection probability can be calculated
exactly and is 
\begin{equation}
P_{\rm d,scint}(I_{\rm T}; S) = \exp\left(-\frac{I_{\rm T}}{S+\Nave} \right ).
\label{eq:pds}
\end{equation}
For a source of {\em observed}, fixed strength $gS$, there exists no 
closed form expression for Eq.~\ref{eq:pd_nonscint}, however.

If there is no signal, $S=0$, the ``detection'' probability 
becomes the ``false-alarm'' probability,  
a measure of how often the intensity will exceed the detection
threshold solely from noise fluctuations.   This false-alarm
probability is 
\begin{equation}
P_{\rm fa}(I_{\rm T}) =  P_{\rm d}(I_{\rm T}; S=0) 
 = \exp\left(-\frac{I_{\rm T}}{\Nave}\right).
\label{eq:pfa}
\end{equation}
By restricting a survey to a certain false-alarm probability, 
one can specify $I_{\rm T}$; 
for example, a false-alarm probability of $10^{-12}$ 
corresponds to $I_{\rm T} = 27.6 \Nave$.
Detection thresholds are set in most surveys so that the ``false-alarm'' rate
yields a small number of spurious ``detections'' over the course of the
survey.

\subsection{Second Order Intensity Statistics}\label{sec:2dpdfs}

The scintillation gain varies on a characteristic time scale, $\Delta t_{\rm d}$,
defined previously.   Later, we will need to consider the joint statistics
of the scintillation gain and the intensity at pairs of times.
The bivariate pdf for the scintillation gain (\cite{goo84}) is:
\begin{equation}
f_{2g}(g_1, g_2; \rho) =  
     \frac{1}{1-\rho}
     \exp\left[ - \frac{ g_1 + g_2}{1 - \rho} \right ]
     I_0\left[ \frac{2\sqrt{g_1 g_2 \rho} } {1 - \rho}\right ],
\label{eq:g_joint}
\end{equation}
where $\rho = \rho(\tau)$ is the temporal correlation function for
the gain $g$, normalized as $\rho(0) = 1$.    This function
is roughly gaussian in form and has a width that is equal to the
characteristic scintillation time of the source.  The scintillation
time, as remarked in \S\ref{sec:tf}, is strongly frequency,
direction, and distance dependent but can be modeled fairly well
(cf. Fig.~\ref{fig:cls_iss}).
For reobservations made promptly such that $\tau\ll \Delta t_{\rm d}$,
$\rho\to 1$ and $f_{2g}$ tends to a one sided exponential function
multiplied by $\delta(g_1 - g_2)$.     For long times between 
observations such that scintillations are independent,
$\rho\to 0$ and $f_{2g}$ becomes the product of the two individual
pdf's for $g_1$ and $g_2$.      

From the bivariate pdf for $g$, we write the bivariate
pdf for the intensity as
\begin{equation}
f_{2I}(I_1, I_2; S,\rho) = \int\int dg_1\, dg_2\, f_{2g}(g_1, g_2; \rho)
    f_I(I_1; g_1 S) f_I(I_2; g_2S).
\label{eq:I_joint}
\end{equation}
As with the scintillation gain, the intensities measured at two times are
more highly correlated for time separations small compared to the scintillation
time; conversely, they are statistically independent if the separations 
exceed the
scintillation time.   The {\em degree\/} of correlation, even at short time
separations, need not be large, however, because this also depends on the
signal to noise ratio.
We write the intensity correlation function as
\begin{equation}
\rho_I(\tau) = \frac{\langle \delta I(t) \delta I(t+\tau)\rangle} 
    {\langle {\delta I^2(t)} \rangle}
\label{eq:rho_I}
\end{equation}
where we define $\delta I$ as the deviation from the mean noise level
\begin{equation}
\delta I(t) \equiv I(t) - \Nave = g(t) S + \delta N + 
    2 \sqrt{g(t) S} {\cal R}(n).
\end{equation}
Defining the noise correlation function as
\begin{equation}
\sigma_N^2\rho_N(\tau) \equiv \langle \delta N(t) \delta N(t+\tau) \rangle
\end{equation}
we obtain, for lags $\tau \ll \Delta t_{\rm d}$,
\begin{equation}
\rho_I(t) \approx 
     \left [
     \frac{g(t) g(t+\tau) S^2 + \Nave^2 \rho_N(\tau) [ 1 + 2Sg(t) / \Nave]}
          { [g(t)S + \Nave ]^2 }
     \right ].
\end{equation}
Note that $\rho_I(0) = \rho_N(0) = 1$.   At a lag $\tau = 0^+$ such 
that the noise has decorrelated completely (i.e., at a lag comparable to
the time needed to calculate a single FFT in a digital Fourier-transform
spectrometer), the intensity correlation is  
\begin{equation}
\rho_I(0^+) \approx \cases{
                \left ( gS / \Nave \right)^2 & $gS \ll \Nave$; \cr
\cr
                1/4                          & $gS = \Nave$; \cr
\cr
                1                            & $gS \gg \Nave$. \cr
}
\label{eq:rhoplus}
\end{equation}
Thus, even though the scintillation gain may be completely correlated
over an observation interval, the intensity need not be.  Moreover, at
rare times, the noise itself can produce a large spike that
decorrelates on a short time scale.

We find that large intensity excursions receive contributions from
both noise and scintillation fluctuations.  Even if the scintillations
are highly correlated at two times, the noise will not be, and the
intensities need not be.  This property has important consequences for
our ability to confirm weak, candidate signals as being real ETI
sources.

It is useful to define a {\em conditional\/} detection probability for
a measurement at a time $\Delta t$ after an initial detection:
\begin{equation}
P_{\rm 2d}(I_T \vert I_1; S, \rho[\Delta t] ) 
      = P\{I_2 > I_T \vert I_1; S, \rho[\Delta t])
      = \int_{I_T}^{\infty} dI_2 \, f_{2I}(I_1, I_2; S, \rho[\Delta t]).
\label{eq:pd_cond}
\end{equation}


\subsection{Intensities at Spatially Separated Sites}\label{sec:sites}

Here we consider joint intensity statistics at two
spatial locations for the purpose of analyzing SETI programs that involve 
simulataneous, dual station observations.    Well separated sites provide
a powerful means for rejecting RFI or instrumental effects that are peculiar
to an individual site (Welch 1983).   However, scintillations and system noise are also
different, in general, at separated sites.    

The intensities measured at two sites at locations $\vec x_{1,2}$ 
will both be of
the form of Eq.~\ref{eq:I} with statistics as described in 
Eq.~\ref{eq:mean}-\ref{eq:pdf_sat}.   We must ask how the different elements of
the intensity are correlated between the two locations.  

The noise is uncorrelated,
except for that resulting from any compact sources 
in the beams of the two telescopes
that would be detected if the two antennas were used as an interferometer.  
At centimetric frequencies and for typical system temperatures, 
any correlated flux
typically amounts to a very small fraction of the total  system noise.  
Henceforth, we ignore any such correlated noise.
 
The scintillation gain $g$ is generally not 100\% correlated between two sites.
In terms of the normalized, temporal
correlation described in \S\ref{sec:2dpdfs},
the spatial correlation is 
$\rho_{\rm s}(\ell = \vperp\tau) = \rho(\ell/\vperp)$, 
and where the characteristic width of of $\rho_{\rm s}$ 
ranges between tens of km for
a strongly scattered source and $>10^4$ km for a weakly 
scattered source
observed at a frequency $\nu = 1$ GHz.   The spatial 
correlation length increases
with frequency as $\ell_{\rm d} \propto \nu^{1.2}$.



For an intrinsic source strength, $S$, the {\em modulated\/} source
strength at the $i^{\rm th}$ site at time $t_{\alpha}$ and frequency
$\nu_{\beta}$ is
\begin{equation}
S_{\alpha\beta i}
 = g(\vec x_i, t_{\alpha}, \nu_{\beta}) S(t_{\alpha}, \nu_{\beta}).
\label{eq:site_s}
\end{equation}
Between two sites, the signal is partially correlated, according to
the value of $\rho_{\rm s}(\ell)$.  The noise at the $i^{\rm th}$ site is
\begin{equation}
N_{\alpha\beta i} = N(\vec x_i, t_{\alpha}, \nu_{\beta})
\label{eq:site_n}
\end{equation}
and is uncorrelated between two different sites 
( $\alpha \ne \alpha^{\prime}$, $\beta \ne \beta^{\prime}$, $i \ne j$)
\begin{equation}
\langle N_{\alpha\beta i} N_{\alpha^{\prime}\beta^{\prime}j} \rangle = 
\langle N(\vec x_i, t_{\alpha}, \nu_{\beta})\rangle 
\langle N(\vec x_i, t_{\alpha}, \nu_{\beta})\rangle. 
\label{eq:cross_n}
\end{equation}

Generally, the cross-correlation of the signal portion of the
intensity involves a multidimensional lag involving spatial
separation, time lag, and frequency separation.  We are interested in
narrowband signals measured at identical times and we will consider
the signals to be intrinsically constant (in time) and deterministic.
For this case, if the only variation in signal between 
two sites is the
spatial difference of the scintillation gain, then:
\begin{equation}
\langle (gS)_{\alpha\beta i} (gS)_{\alpha\beta j} \rangle 
 = \langle g(\vec x_i, t_{\alpha}, \nu_{\beta})
        g(\vec x_j, t_{\alpha}, \nu_{\beta})\rangle 
        S^2(\nu_{\beta}) 
 = [1 + \rho_{\rm s}(\vert\vec x_i - \vec x_j\vert)] S^2(\nu_{\beta}).
\label{eq:cross_s}
\end{equation}
In practice, there is a Doppler shift of the signal 
between the two sites that must be accounted for and which
provides useful means for assessing the celestial nature
of a source.  But as long as the Doppler shift is less
than the characteristic bandwidth of the scintillations,
Eq.~\ref{eq:cross_s} still applies. Larger Doppler
shifts can be described by a similar equation,
but with an effective $\rho_s$ that is a combination
of spatial and frequency correlations.




\newpage

\begin{figure}
\caption[Contours of Constant Scintillation Time]
{{\sl Top:} Contours of the scintillation time scale as viewed looking down on
the plane of the Galaxy.  The $+$ indicates the Galactic center and
the dashed circles have radii of 5 and 10~kpc.  The offset figure
shows the assumed locations of the spiral arms.  The scintillation
times refer to an observation frequency of 1.42~GHz and for directions
at zero Galactic latitude.  The scintillation time scales with
frequency as $\nu^{1.2}$.  
{\sl Bottom:}
Contours of the scintillation bandwidth at 1.42~GHz.
The bandwidth scales with frequency as $\nu^{4.4}$.}
\label{fig:cls_iss}
\end{figure}


\begin{figure}
\caption[Detection Probabilities]
{Detection probabilities vs normalized source strength $S/I_{\rm T}$
for the case where the source does not scintillate (dashed line) and
for the case where it does scintillate in the strong scattering regime
(solid line).  }
\label{fig:cls_pd}
\end{figure}


\begin{figure}
\caption[Conditional Detection Probability]
{Detection probability for a second detection above a threshold
$\eta_{\rm T} = I_{\rm T}/\Nave$ given an initial detection with
intensity $\eta_1 = I_1/\Nave$.  Curves are plotted as 
a function of
$\rho_g$, which may be either the spatial or
temporal correlation function of the scintillation gain.
Curves are labelled with the {\em
intrinsic\/} source strength $\zeta = S/\Nave$, i.e., the source
strength in the absence of scintillations.
{\it a}) $\eta_1 = 32$, $\eta_{\rm T} = 20$, typical META parameters;
{\it b}) $\eta_1 = 32$, $\eta_{\rm T} = 10$, recommended
reobservation threshold for future META-like surveys;
{\it c}) $\eta_1 = 16$, $\eta_{\rm T} = 16$, 
typical dual-site META-META~II parameters;
{\it d}) $\eta_1 = 10$, $\eta_{\rm T} = 10$, recommended thresholds
for future dual-site observations.  }
\label{fig:cls_p2cond}
\end{figure}

\begin{figure}
\caption[Time Series of Signals for the Three Models]
{Representative time series for the three models for META
events; the dashed lines indicate a detection threshold: 
{\it a}) Noise only.
{\it b}) A scintillating source with amplitude $gS = 5\Nave$ combined with noise. 
    It is assumed that the scintillation gain is constant over
    the time series.
{\it c}) A scintillating source with time-varying scintillation gain, $g(t)$,
    combined with noise.
{\it d}) RFI pulse combined with noise.   }
\label{fig:cls_ts}
\end{figure}

\begin{figure}
\caption[Contours of A Posteriori PDF]
{The joint {\em a posteriori\/}  pdf for the scintillation gain $g_{\rm c}$ and 
intrinsic signal strength $\zeta = S/\Nave$.  Contours are at 
95\%, 90\%, 50\%, 10\%, 1\%, and 0.1\% of the peak.  The number
of reobservations is assumed to be 256.
{\it a}) For a candidate with $\eta_{\rm c} = 32$ and a reobservation threshold
of $\eta_{\rm T}^{(\reobs)} = 20$, typical of the META survey.
{\it b}) For $\eta_{\rm c} = 100$ and $\eta_{\rm T}^{(\reobs)} = 20$.
{\it c}) For $\eta_{\rm c} = 32$ and $\eta_{\rm T}^{(\reobs)} = 10$.
{\it d}) For $\eta_{\rm c} = 100$ and $\eta_{\rm T}^{(\reobs)} = 10$.
}
\label{fig:cls_eventpdf}
\end{figure}

\begin{figure}
\caption[Best fit values for $g_{\rm c}$, $\zeta$, and their product
$g_{\rm c}\zeta$.]
{Best fit values for $g_{\rm c}$, $\zeta$, and their product $g_{\rm
c}\zeta$ for an initial
candidate detection plotted against the number of reobservations.
It is assumed that the scintillation gain $g$ is uncorrelated between
reobservations and between any reobservation
and the original survey detection.   Results are shown for a survey
amplitude $\eta_{\rm c} = 32$ and for three values of the reobservation
threshold, $\eta_{\rm T}^{(\reobs)} = 5$, 15, and 20.
{\it a}) $\zeta$ vs. $\kru$. 
{\it b}) $g_{\rm c}$ vs. $\kru$. 
{\it c}) $g_{\rm c}\zeta$ vs. $\kru$. 
}
\label{fig:cls_gz_rho0}
\end{figure}

\begin{figure}
\caption[Critical Number of Reobservations]
{[Top] The critical number of reobservations $\krucrit$ plotted against
reobservation threshold $\eta_{\rm T}^{(\reobs)}$.  $\krucrit$ is the number of
reobservations needed for a given reobservation threshold
$\eta_{\rm T}^{(\reobs)}$ in order to render Model~II highly
improbable.  The curves are labelled by the intensity of the candidate
signal, $\eta_{\rm c}$, at the initial detection.
[Bottom]  The number of false alarm detections expected (per frequency channel)
in $\krucrit$ reobservations.   The dashed line designates one false alarm
per frequency channel, an unacceptably large number.  
}
\label{fig:cls_krucrit}
\end{figure}

\begin{figure}
\caption[Best fit values for $g_{\rm c}$, $\zeta$, and $g_{\rm c}\zeta$ for $\krc \ne 0$.]
{Similar to Fig.~\protect\ref{fig:cls_gz_rho0}, 
but where correlated reobservations, $\krc$, 
are included in the analysis.
{\it a}) Candidate intensity $\eta_{\rm c} = 32$ and reobservation
threshold $\eta_{\rm T}^{(\reobs)} = 20$.
{\it b}) $\eta_{\rm c} = 32$ and $\eta_{\rm T}^{(\reobs)} = 10$.
{\it c}) $\eta_{\rm c} = 32$ and $\eta_{\rm T}^{(\reobs)} = 5$.
}
\label{fig:cls_gz_rho100}
\end{figure}

\begin{figure}
\caption[Critical Number of Reobservations]
{As for Fig.~\protect\ref{fig:cls_krucrit} but for 
{\em correlated\/}  reobservations $\krccrit$.
}
\label{fig:cls_krccrit}
\end{figure}


\begin{figure}
\caption[Model~II and III Likelihood Function for META Candidates]
{Contours of the logarithm of the survey likelihood functions for the
META candidates.  Contours are at 95\%, 90\%, 50\%, 10\%, 1\%, and
0.1\% of the peak in each panel.
{\it a}) The survey likelihood function, including reobservations, of
	the four META candidates at 1420~MHz assuming Model~II, namely
	a population of standard candle transmitters in the Galactic
	disk which are modulated by scintillation.
{\it b}) As for ({\it a}), but for the seven 2840~MHz candidates.
{\it c}) As for ({\it a}), but for Model~III, namely RFI.
{\it d}) As for ({\it b}), but for Model~III, namely RFI.
}
\label{fig:meta_models}
\end{figure}

\begin{figure}
\caption[Sky Fraction vs. Mean Free Path]
{The fraction of the sky having ETI sources in the telescope beam,
$\epsilon_{\rm sky}$, as a function of the mean free path for a beam
to intersect an ETI source.  We assume a uniform galactic disk
of radius 15 kpc and thickness 0.2 kpc.}
\label{fig:cls_epssky}
\end{figure}

\begin{figure}
\caption[Number of Sources vs. Survey Fraction]
{
The number of sources $N_{\rm D}$ plotted against survey fraction
$\epsilon_{\rm su}$.  The solid curve is the estimate for
$N_{\rm D}$ obtained from Eq.~\ref{eq:nd1} while the dashed line
results from Eq.~\ref{eq:nd2}.    
}
\label{fig:cls_solutions}
\end{figure}


\newpage

\setcounter{table}{0}
\begin{planotable}{llll}
\tablewidth{400pt}
\tablecaption{SETI PROGRAM PARAMETERS
\label{tab:surveys}}
\tablehead{\colhead{Parameter}&\colhead{META}& \colhead{META II}& \colhead{SERENDIP III} }
\startdata
$N_{\nu}$ 		&  2  		&  1  		&  5 \nl
$N_{\rm frames}$	& 3		& 3		& 1 \nl
$N_{\rm ch}$		& $2^{23}$	& $2^{23}$	& $2^{22}$ \nl
$N_{\rm pol}$		& 2		& 2		& 1 \nl
$N_{\rm sky}$		& $10^{5.2}$	& $10^{5.0}$	& $10^{5.8}$ \nl
$K_{\rm s}$		& 4		& 1		& 2.3 (ave) \nl
$N_{\rm trials}$	& $10^{13.8}$	& $10^{13.2}$	& $10^{13.9}$ \nl
Beam size (deg) 	& 0.5 (1.42 GHz)& 0.5 (1.42 GHz)& 0.15 (0.42 GHz) \nl
                	& 0.25 (2.84 GHz) \nl
$\Delta\nu_{\rm ch}$ (Hz)& 0.05    	&  0.05         & 0.6 \nl
$\Delta\nu_{\rm total}$ (MHz) & 0.4 	&  0.4 		& 12  \nl
dwell time (s)		& 20 		& 20		& 1.7 \nl  
\multicolumn{4}{l}{Thresholds:}\nl
\quad survey     	& 31.7$\Nave$	& 24$\Nave$     & 15$\Nave$ \nl
\quad reobserve  	& 20$\Nave$	& \nodata       & \nodata \nl
\multicolumn{4}{l}{Sensitivities: (EIRP = effective isotropic radiation power.)}\nl
\quad EIRP (W m$^{-2}$) & $10^{-22.8}$	&		& $10^{-24.6}$
\nl 
\enddata
\end{planotable}

\newpage
\begin{planotable}{rrrr}
\tablewidth{210pt}
\tablecaption{Bayesian Maxima for META}
\label{tab:posterior}
\tablehead{\colhead{$\eta_{\rm c}$}&\colhead{$\eta_{\rm T}$}&
	\colhead{$g_{\rm c}$}&\colhead{$\zeta$}}
\startdata
32 & 20 & 6.2 & 2.6 \nl
100 & 20 & 16.3 & 3.9 \nl
\nl
\nl
32 & 10 & $< 10^{-3}$ & $< 10^{-1}$ \nl
100 & 10 & 22.0 & 2.0 \nl
\nl
\tableline
\enddata
\end{planotable}
\newpage

\newpage
\begin{planotable}{lccr}
\label{tab:metasurvey}
\tablewidth{210pt}
\tablecaption{Likelihood Maxima for META}
\tablehead{\colhead{$\nu$}&\colhead{max($\Lambda$)}&
	\colhead{$\zeta_0\epsilon_{\survey}$}&\colhead{$\zeta_1$} \\
\colhead{(MHz)}&\colhead{ }&\colhead{ }&\colhead{ }}
\startdata
\multicolumn{4}{c}{Model II: Survey}\nl
1420 & \phantom{1}-61.6 & $10^{-11.9}$ & $10^{2.78}$ \nl
2840 & -107\phantom{.6} & $10^{-11.6}$ & $10^{3.54}$ \nl
\nl
\tableline
\nl
\multicolumn{4}{c}{Model II: Survey+Reobservations} \nl
1420 & \phantom{1}-61.6 & $10^{-11.9}$ & $10^{2.78}$ \nl
2840 & -107\phantom{.6} & $10^{-11.6}$ & $10^{3.54}$ \nl
\nl
\tableline
\tableline
\nl
\multicolumn{4}{c}{Model III: Survey} \nl
1420 & \phantom{1}-61.3 & $10^{-13.6}$ & $10^{2.42}$ \nl
2840 & -108\phantom{.3} & $10^{-13.2}$ & $10^{2.91}$ \nl
\nl
\tableline
\nl
\multicolumn{4}{c}{Model III: Survey+Reobservations} \nl
1420 & \phantom{1}-61.3 & $10^{-13.6}$ & $10^{2.42}$ \nl
2840 & -108\phantom{.3} & $10^{-13.2}$ & $10^{2.91}$ \nl
\enddata
\end{planotable}

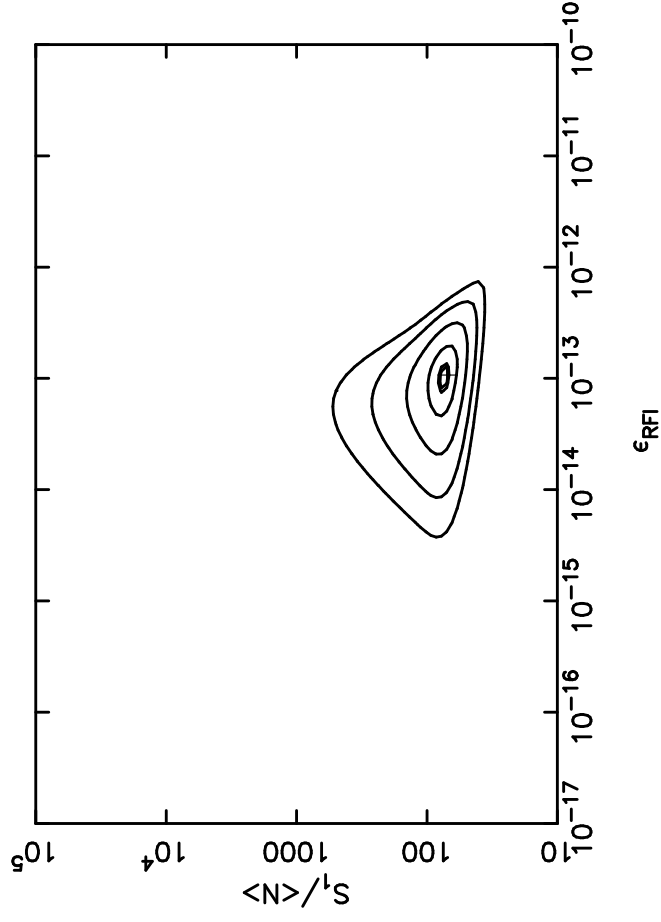
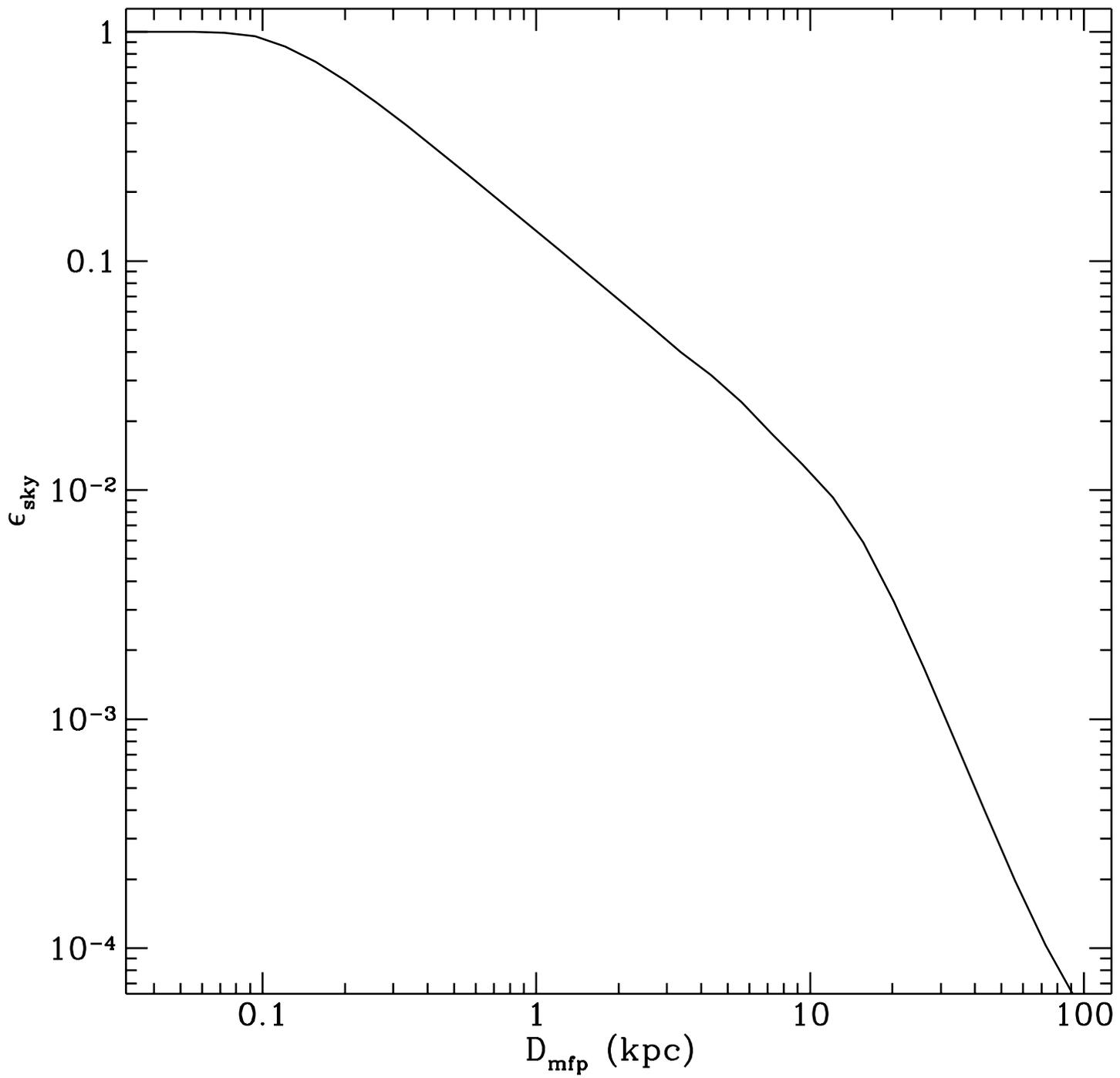
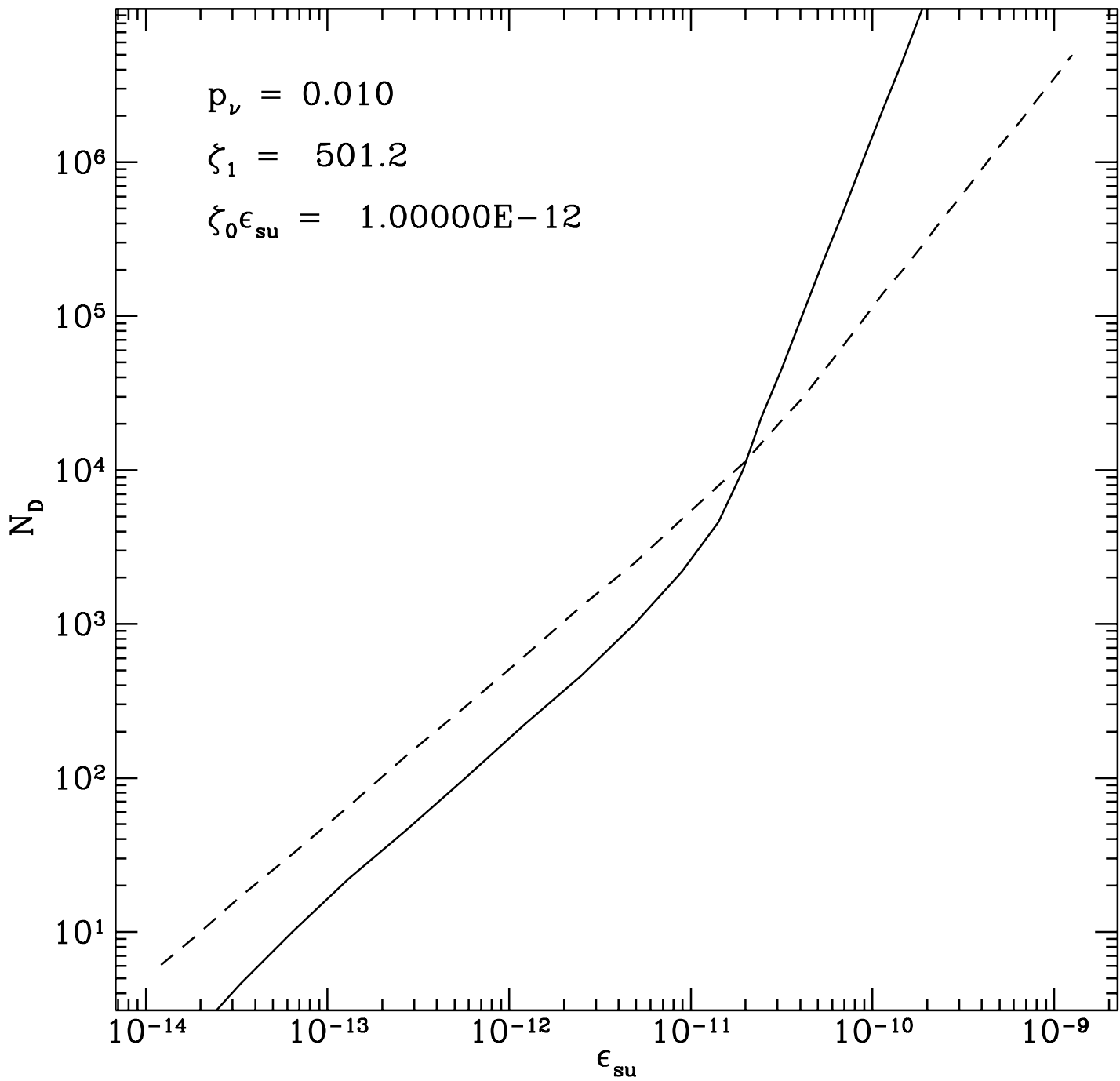
\end{document}